\documentclass[journal]{IEEEtran}
\usepackage[utf8]{inputenc}
\usepackage{cite}
\usepackage{amsmath,amssymb,amsfonts}
\usepackage{algorithmic}
\usepackage{graphicx}
\usepackage{textcomp}
\usepackage{xcolor}
\usepackage{color,soul}
\def\BibTeX{{\rm B\kern-.05em{\sc i\kern-.025em b}\kern-.08em
    T\kern-.1667em\lower.7ex\hbox{E}\kern-.125emX}}
    
\usepackage[ruled,vlined]{algorithm2e}
\usepackage{listings}
\usepackage[colorinlistoftodos]{todonotes} 
\usepackage{import}
\usepackage{array}

\definecolor{mGreen}{rgb}{0,0.6,0}
\definecolor{mGray}{rgb}{0.5,0.5,0.5}
\definecolor{mPurple}{rgb}{0.58,0,0.82}
\definecolor{backgroundColour}{rgb}{0.95,0.95,0.92}
\definecolor{mygreen}{rgb}{0,0.6,0}
\definecolor{mygray}{rgb}{0.5,0.5,0.5}
\definecolor{mymauve}{rgb}{0.58,0,0.82}

\lstdefinestyle{CStyle}{
  backgroundcolor=\color{backgroundColour}, 
  breakatwhitespace=false,         
  breaklines=true,                 
  captionpos=b,                    
  deletekeywords={...},            
  escapeinside={\%*}{*)},          
  extendedchars=true,              
  keepspaces=true,                 
  morekeywords={*,...},            
  numbers=left,                    
  numbersep=5pt,                   
  numberstyle=\tiny\color{mygray}, 
  rulecolor=\color{black},         
  showspaces=false,                
  showtabs=false,                  
  tabsize=2,	                   
  title=\lstname,                   
  belowcaptionskip=0pt,
  belowskip=-0.8 \baselineskip,
  xleftmargin=\parindent,
  language=C,
  showstringspaces=false,
  basicstyle=\footnotesize\ttfamily,
  keywordstyle=\bfseries\color{green!40!black},
  commentstyle=\itshape\color{purple!40!black},
  identifierstyle=\color{blue},
  stringstyle=\color{orange}
}

\usepackage{subcaption}

\usepackage[colorinlistoftodos]{todonotes} 

\newcommand{\intset}[1]{\ensuremath{\left[\!\left[#1\right]\!\right]}}
\newcommand{\keymsg}[1]{\textbf{#1}}
\newtheorem{proposition}{Proposition}

\title{Increasing FPGA Accelerators Memory Bandwidth with a Burst-Friendly Memory Layout}
\author{\IEEEauthorblockN{Corentin Ferry\IEEEauthorrefmark{1}\IEEEauthorrefmark{2},
Tomofumi Yuki\IEEEauthorrefmark{1},
Steven Derrien\IEEEauthorrefmark{1} and
Sanjay Rajopadhye\IEEEauthorrefmark{2} 
}\\
\IEEEauthorblockA{\IEEEauthorrefmark{1}Univ Rennes, CNRS, IRISA, Inria. Rennes, France.}\\
\IEEEauthorblockA{\IEEEauthorrefmark{2}\textit{Department of Computer Science}, \textit{Colorado State University}. Fort Collins, Colorado, USA}
}

\date{September 2021}

\begin{document}

\maketitle

\begin{abstract}
Offloading compute-intensive kernels to hardware accelerators relies on the large degree of parallelism offered by these platforms. However, the effective bandwidth of the memory interface often causes a bottleneck, hindering the accelerator's effective performance. Techniques enabling data reuse, such as tiling, lower the pressure on memory traffic but still often leave the accelerators I/O-bound. A further increase in effective bandwidth is possible by using burst rather than element-wise accesses, provided the data is contiguous in memory.

In this paper, we propose a memory allocation technique, and provide a proof-of-concept source-to-source compiler pass, that enables such burst transfers by modifying the data layout in external memory. We assess how this technique pushes up the memory throughput, leaving room for exploiting additional parallelism, for a minimal logic overhead.
\end{abstract}

\section{Introduction}

Hardware acceleration of compute-intensive mathematical kernels is one of the most efficient ways to improve their performance. Compute-intensive algorithms, such as image processing algorithms or neural networks, can benefit greatly from the use of the massive parallelism and low latency of application-specific hardware. There are however significant hurdles preventing widespread use of such hardware: aside from the manufacturing cost of an application-specific integrated circuit (ASIC), which a programmable chip such as a field-programmable gate array (FPGA) can mitigate to some extent, the development effort needed to get a decently performing accelerator is orders of magnitude greater than developing a GPU version of the same algorithm.

\begin{figure}
    \centering
    \includegraphics[width=\columnwidth]{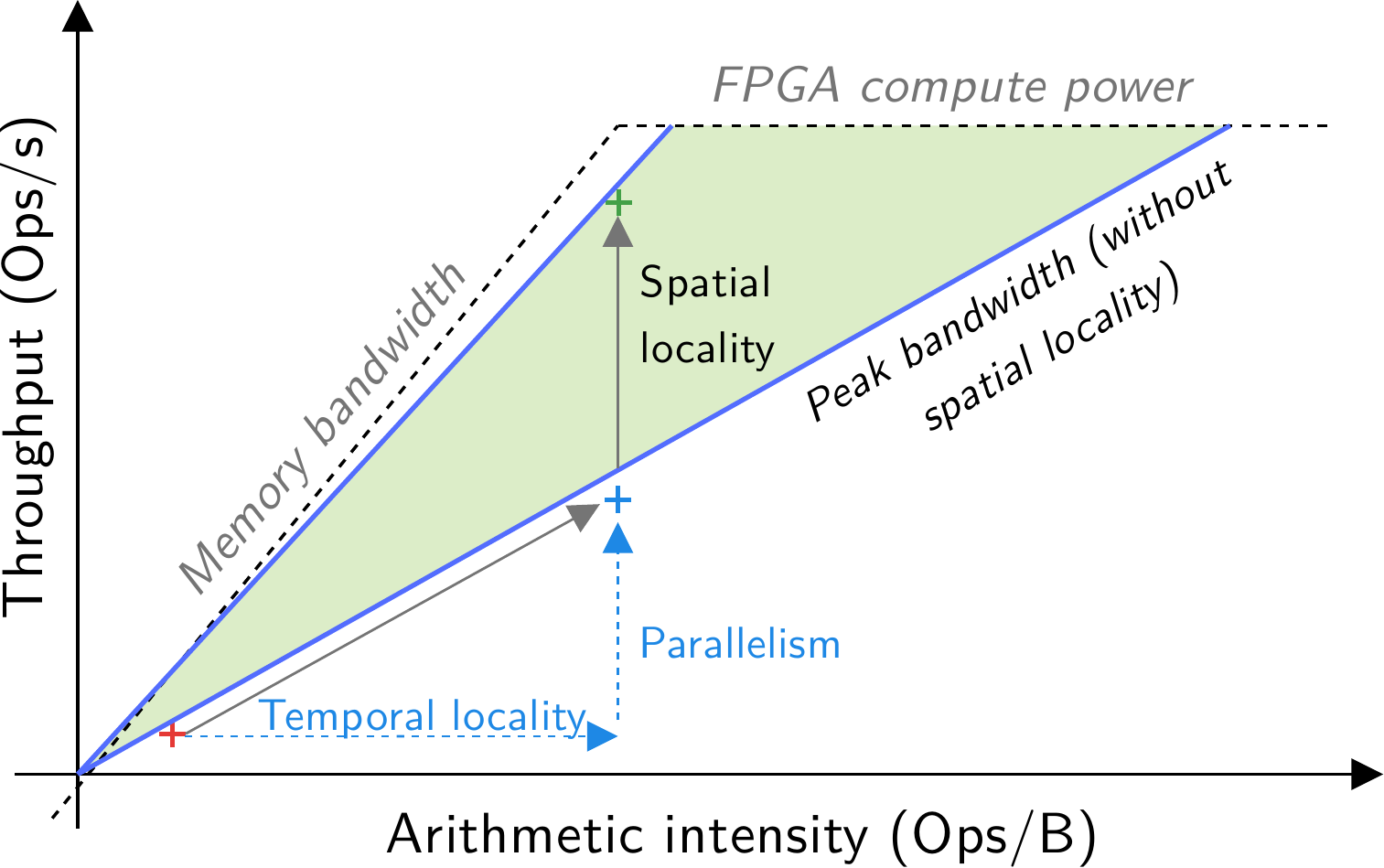}
    \caption{Improving both temporal and spatial locality is needed to get a higher compute throughput. Limited amounts of on-chip memory means that temporal locality cannot be improved beyond a certain limit, therefore using spatial locality to increase memory bandwidth becomes necessary.}
    \label{fig:roofline}
\end{figure}

Design automation tools have the ability to create massively-parallel accelerators. The result has so much processing power that memory accesses are not fast enough to feed the accelerator, which is then called \emph{memory-bound} (the opposite situation, where the processing power is the bottleneck, being \emph{compute-bound}). Cong et al. \cite{Cong_2018} emphasize that FPGAs have as much computing power as GPUs, but not the memory bandwidth. One way to address this \emph{memory wall} \cite{Williams_2009} is to increase the \emph{arithmetic intensity}, by either computing more for the same memory traffic or transferring less for the same amount of computations. Even so, memory-bound accelerators are still commonplace. Random memory accesses are among the primary causes of the memory wall, that lowers the effective bandwidth as illustrated by Figure \ref{fig:roofline}.

Favoring burst accesses over random accesses is a key to increasing the the effective bandwidth, and therefore the performance of the accelerator. Doing so implies transforming the accelerator's memory access pattern; to be profitable, such a transformation must not degrade the on-chip compute power. 

In this work, we propose a transformation that improves spatial locality of programs with uniform memory accesses, and integrate it into a polyhedral compiler flow. Broad classes of actively studied programs fall into this category, notably convolutions, FDTD or matrix multiplication. The idea is to transform the off-chip data layout and get as many contiguous off-chip accesses as possible, while preserving the compute-oriented on-chip optimizations.

The contributions of this work are:
\begin{itemize}
  \item A memory layout and access pattern that favor burst accesses,
  \item A proof-of-concept compiler pass that automatically applies this memory layout,
  \item An evaluation of the performance of this layout, that shows it can fully utilize the available bandwidth, without a significant increase in area.
\end{itemize}

Our work relies on well-known techniques, such as iteration space tiling \cite{Wolf_1991, Irigoin_1988}, and tools, such as the Integer Set Library (ISL) \cite{Verdoolaege_2010} to represent polyhedral sets and Pluto \cite{Bondhugula_2008} to apply iteration space tiling and generate a schedule.

This paper is organized as follows: first, it introduces the notion of data locality and on-chip parallelism-enabling techniques onto which we rely (Sections  \ref{sec:bground} and \ref{sec:relwork}); then, it describes the construction of our burst-friendly off-chip layout (Section \ref{sec:cfa}). Finally, it provides an evaluation and comparison with the state of the art (Section \ref{sec:eval}).

\section{Background}
\label{sec:bground}

This section explains the issue of memory bandwidth under-utilization, and the challenges that one faces to improve its use.

\subsection{Increasing compute performance makes the accelerator memory-bound}
\keymsg{Fine-tuning programs to use architectural optimizations (or architectural features on CPUs/GPUs) makes the design memory-bound.}

In order to get the maximum performance with a given architecture, a program needs to fully exploit its hardware compute resources. This typically means using parallel execution units. Vector units are designed specifically for this purpose on CPUs and GPUs, with dedicated register files. FPGA and ASIC designers have to build their own on-chip architecture where processing units and memories can be actively used at all times.

High-level synthesis tools also have the ability to perform loop unrolling, akin to automatic loop vectorization for CPUs which has been part of state-of-the-art compilers for years \cite{Bik_2002}, in order to exhibit more parallel computations. Loop pipelining,
also available as a loop transformation in HLS tools, increases the compute throughput by ensuring all functional units are kept busy at all times. Combinations of these techniques \cite{Pouchet_2013}, \cite{Zhang_2015} yield drastic performance increase over non-parallel implementations.

Increasing parallelism often also increases memory traffic: to perform more operations at a time, an accelerator may need more data. Increasing parallelism without increasing arithmetic intensity makes the accelerator memory-bound, thus there is a need for memory optimizations such as data compression \cite{Nakahara_2020}.

\subsection{Improving locality increases memory access performance}
\keymsg{To get a better performance, we need to reduce pressure on memory, and to do that, we need to improve both temporal and spatial locality.}

A general rule that applies to all memory systems regardless of their hierarchy is that the further data is from the compute engine, the longer it takes to access it. Faster memory cells such as registers, close to the compute engines, are expensive and therefore scarce. Numerous program transformations seek to minimize the latency incurred by memory accesses, by improving the program's behavior in two ways:
\begin{itemize}
  \item Minimizing cache misses. This means using data available at closer cache levels in priority, and computing as much as possible with it. As caches contain the most recently accessed data, this is called improving \textbf{temporal locality}.
  \item Accessing data in the same order as it is stored in memory, thereby exploiting the burst facilities offered by the memory controller and DRAM chips, resulting in a higher throughput. This is called improving \textbf{spatial locality}.
\end{itemize}

The \emph{roofline model} \cite{Williams_2009}, illustrated on Figure \ref{fig:roofline} is a visual way to see the effects of locality improvements. The throughput a system can reach is bound when either a \emph{memory roofline} or a \emph{compute roofline} is hit. The former case incurs an under-utilization of the on-chip parallel compute resources, which is undesirable, where the latter incurs idle memory bus time, which is good in regards to energy, and means no further memory improvements are needed. Figure \ref{fig:roofline} shows that improving both locality metrics is needed to go from the former to the latter case. The next subsections will go over locality-improving techniques and why further spatial locality improvement is needed.

\subsection{Tiling loops to improve temporal locality}
\keymsg{Temporal locality is obtained by transforming loops, in particular tiling them.}

A common transformation to improve temporal locality is called loop tiling. Its principle is to break the workload of a loop (its \emph{iteration space}) into smaller, single-shaped packets called \emph{tiles}. Such tiles are small enough so that the amount of data required for them to execute fits in a local memory or cache. Tiling therefore increases temporal locality and arithmetic intensity. 

The impact of tiling on performance is such that finding the best tile size and shape for specific classes of programs, such as matrix multiplication \cite{Clint_2001} or convolution \cite{Zhang_2015}, is a trending topic in compiler research.

Given the performance gain permitted by tiling, our work will focus on improving spatial locality specifically for tiled loops.

\subsection{Tiling enables overlapping communications from execution}
\keymsg{In order to use two data layouts and gain transfer performance without losing compute power, we have to separate communications and computations. It is always possible to do so with tiled loops due to atomicity of a tile.}

To sustain a high performance, all parts of the accelerator should stay active at all times, whether they are dedicated to computation or communication. Thanks to the atomic nature of tiles, it is possible by definition to read a tile's input ahead of execution, and write its output to memory after execution. It is therefore possible to split the workload into three tasks, each one processing a different tile: while a tile of iterations is being executed, the accelerator can prepare the execution of the next tile by reading its input data, and write the results of the previous tile. 

\begin{figure}
\centering
\includegraphics[width=\columnwidth]{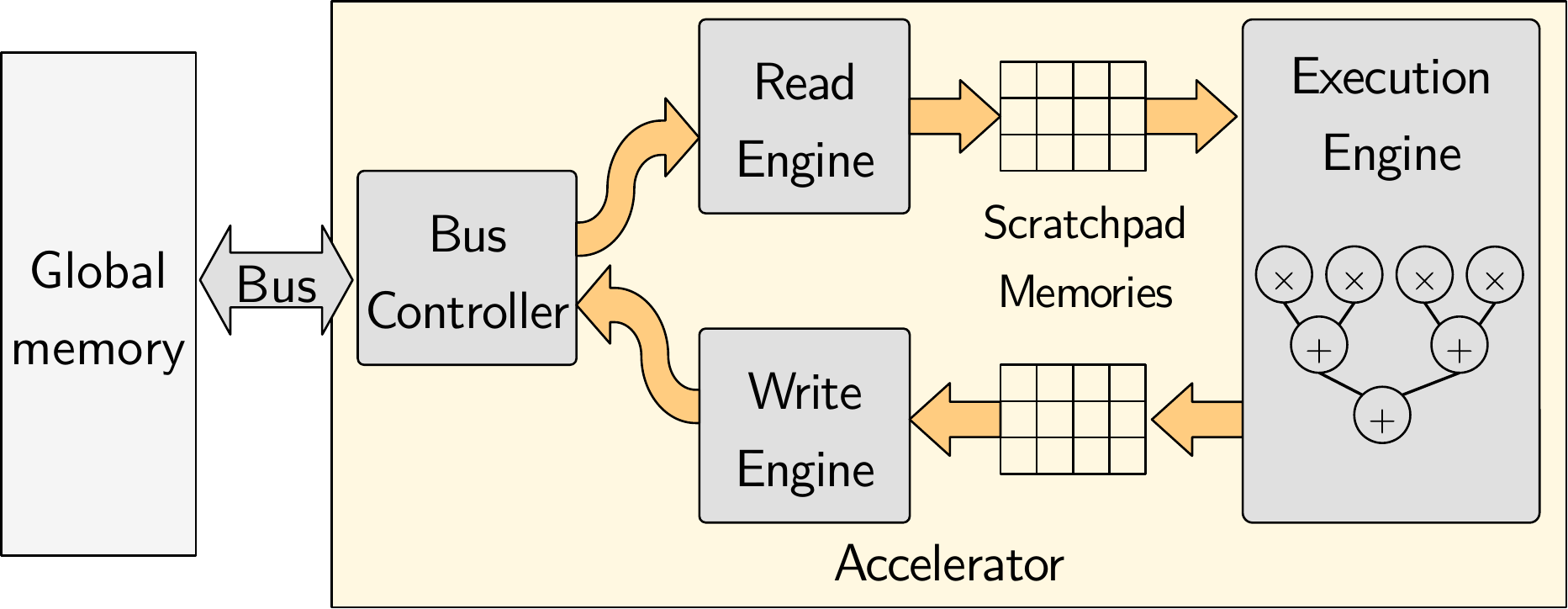}
\caption{Task-level structure of an accelerator: read, execution and writeback happen separately. \emph{On-chip memory accesses are random, while off-chip accesses are contiguous.}}
\label{fig:CFAAcceleratorStructure}
\end{figure}

The usual structure of a hardware accelerator for a tiled loop, shown in Figure \ref{fig:CFAAcceleratorStructure}, follows this read-execute-write template. Communications are overlapped with computation in the form of a \emph{task-level pipeline}, that HLS tools such as Vivado and Catapult provide ways to achieve automatically. 

A task-level pipeline significantly increases the overall throughput: instead of processing a new tile after each one has completed the read-execute-write tasks, a new tile starts each time the readback of a tile completes, and at the same time, the execution and writeback of two other tiles complete.

\subsection{Full bandwidth usage requires high spatial locality}
\keymsg{Tiling alone does not bring spatial locality. The access pattern has to be transformed to achieve that.}

Loop tiling, in general, increases temporal locality by having a tile's footprint fit in a cache or local memory. Traffic from / to global memory may still have a low throughput, keeping the program memory-bound. The memory access latency, and therefore the memory throughput, depends on the sequence of accessed addresses, which is called the memory access pattern.

The access pattern impacts a memory system's performance at several levels. At the DRAM chip level, switching rows and banks incurs a certain latency; reading physically contiguous data at consecutive addresses therefore performs better than accessing random locations. At the controller level, each transaction incurs an incompressible latency. Random accesses require one transaction each; burst accesses, which are reads or writes to several consecutive addresses, take in a single transaction. Making use of this feature gives a lower latency requires a contiguous access pattern. 

The more contiguous accesses an access pattern has, the better its memory throughput is. Transforming a pattern this way is said to improve its spatial locality.

\subsection{Spatial locality should be applied on flow-in and flow-out sets} \label{flow-in-out-introd}
\keymsg{The flow-in/out sets are the ones which need to be transformed to exhibit spatial locality.}

Part of the data produced by a tile will be used later on by another tile. Given that tile size is usually determined to fill up on-chip memories, there is not enough on-chip memory to hold a tile's produced data for later use. Therefore this data needs to be stored in global memory. This specific set of data is called the \emph{flow-out} set of the tile. Likewise, the \emph{flow-in} set of a tile is those data produced earlier that need to be brought on chip before they are used.

Application-specific hardware accelerators usually feature scratchpad memories instead of caches. Copying the data in and out of scratchpads is up to the accelerator. The atomic nature of tiles allows splitting an accelerator's work into three discrete steps: read the flow-in data, perform the actual execution, and write the flow-out data to memory.

The flow-in and flow-out sets being the ones generating off-chip memory traffic, they are the ones for which we need to optimize the transfer from and to global memory.

\subsection{Representing loops using the polyhedral model} 
\keymsg{We need a set representation of loops that supports transformations; the polyhedral model is a fit.}

Loop transformations are more easily expressed as closed-form mappings of vector spaces, than as operations on loop nests. The usual program representation used for tiling is a compact, closed-form representation of a its control and data flows, called the polyhedral model.

Loops are mapped to a vector space called \emph{iteration space} where each loop index is a dimension of this space. An affine function called \emph{schedule} maps each iteration to an integer, giving an order of execution of iterations. 

Constraints on the schedule are set by the \emph{dependence pattern}, which is a set of affine functions mapping an iteration to those that it uses the result of: any iteration must be scheduled after those it depends on. This work focuses on programs with uniform dependencies; those are of the form $\vec{x} \mapsto \vec{x} + \vec{b}$ where $\vec{b}$ is a constant vector.

In the scope of this work, we will be performing affine operations on iteration vectors such as projection; it should be noted that tiling can be seen as the composition of affine operations on the iteration space.

\subsection{Representing memory accesses using affine functions}
\keymsg{Physical memory access functions can be decomposed into array access function + memory layout.}

\begin{figure}
\centering
\includegraphics[width=\columnwidth]{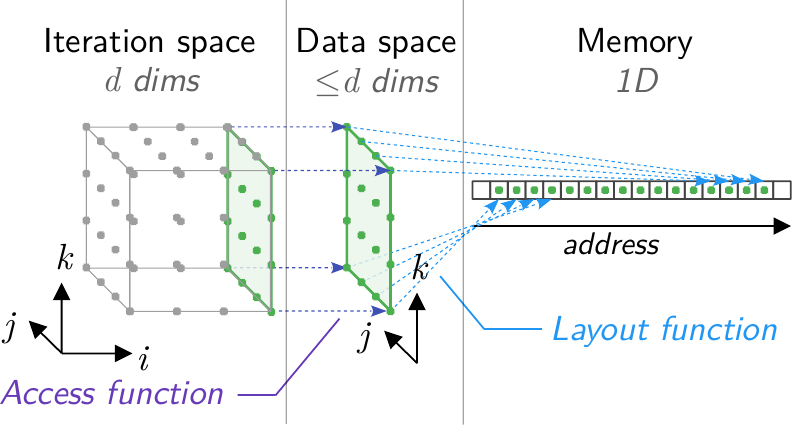}
\caption{Polyehdral iteration-to-memory flow: each memory access is the combination of an array access function and the layout function.}
\label{fig:iteration2memflow}
\end{figure}

Every memory access performed by a program is made at an address that is computed at runtime from the loop indices using some \emph{array access function}. Most programming languages support multi-dimensional arrays, making it possible to use multiple indices to access memory cells; however, memory is one-dimensional, so each multi-dimensional array must be mapped to a one-dimensional array in memory. Such a mapping is called a \emph{memory layout}. Figure \ref{fig:iteration2memflow} shows the decomposition of memory accesses into the composition of array access functions and memory layout.

In the polyhedral model, access and memory layout functions are affine functions that respectively go from the iteration space to a data space, and from a data space to memory. There is one data space per array in the program, that mirrors an array, e.g. the array \texttt{int A[200][100]} is mapped to the set $\left\lbrace A[i, j] : 0 \leqslant i < 200\text{ and }0 \leqslant j < 100\right\rbrace$.

\section{Related work}
\label{sec:relwork}

The high-level issue our work is addressing is the memory wall that appears on Figure \ref{fig:roofline}, which is hindering the overall system performance. Techniques that aim to relieve it can be sorted into three main categories: those that increase the arithmetic intensity (right-wards on a roofline graph), those that increase performance (upwards on a roofline graph), and those that increase the effective bandwidth.

\subsection{Increasing the arithmetic intensity}
\keymsg{Increasing the AI reduces the pressure on memory in terms of volume of communication, be it with spatial locality / data reuse or with approximation.}

Increasing the arithmetic intensity, per definition, means reducing the volume of data transferred per arithmetic operation. The main way to do that is to re-use the data already locally present on-chip or in a close level of the memory hierarchy.

\textbf{Playing on the program's schedule} to favor data reuse reduces the need for off-chip transfers. Loop tiling \cite{Irigoin_1988, Wolf_1991} is the main technique in this aim. The primary target of tiling was to reduce the cache miss rate of CPUs, that have a cache hierarchy, but the technique also applies to software-programmable accelerators such as GPUs \cite{Holewinski_2012}
, and thanks to high-level synthesis tools, it also applies to application-specific hardware accelerators as well \cite{Pouchet_2013}

\textbf{Sharing on-chip resources to maximize their usage}: On hardware accelerators, fine-tuning of on-chip memory allocation may eliminate off-chip accesses. Memory cells may be allocated multiple times, shared across tasks that do not interfere \cite{Wei_2019}, increasing the usefulness of each memory cell, thereby reducing the amount of off-chip traffic.

\subsection{Increasing the effective bandwidth}
The root of the memory bandwidth issue is found in the massive parallelism the hardware can provide. When it is fully used, the memory latency may be higher than the compute latency, due to how the memory subsystem is designed and used. In this case, the design is said to be \emph{memory-bound}. Such a limitation may be caused, for instance, by port contention, cache conflict misses, scalar accesses to global memory. All of these issues find their root in a sub-optimal memory access pattern, a sub-optimal physical placement (layout) of the data, or both. 

Program and data transformations of several kinds can be used to adapt the access pattern to the data layout, or adapt the data layout itself to the program; proper use of the memory and its access interfaces result in an increase of the memory bandwidth to a value closer to the nominal memory chip bandwidth. Further increase of the \emph{effective bandwidth}, which is the amount of useful data transferred per unit of time, may be achieved using data compression techniques. 

Our wor
is positioned among these techniques, but it is also essential that temporal locality and on-chip performance optimizations are applied to exhibit sufficient parallelism and create the demand for bandwidth.

\subsubsection{Optimizing memory access pattern and layout}
\keymsg{Making an optimal use of bandwidth means have an on-chip data layout and access pattern free of port contention, and exhibit contiguity in off-chip accesses.}

Sub-optimal usage of memory bandwidth may be due to where the data is located in memory and the schedule of access requests. These can be tuned to alleviate certain issues they are at the root of.

\paragraph{Eliminating access conflicts} Access conflicts may occur in all memory architecture, when one wants to use a physical port or memory cell to convey multiple data at the same time. Mapping the data to other cells or addresses will resolve the conflicts. On FPGA chips, parallel memory access patterns may incur port contention. Such a phenomenon is common with on-chip memories, but may happen on multiple-port memory architectures such as high-bandwidth memory (HBM). Bank partitioning \cite{Cong_2011} is a key and widely used technique to bank conflict prevention, and is available in commercial high-level synthesis tools. Conflicts also happen in set-associative caches when multiple addresses share the same set and therefore the same physical memory cells in the cache. Appropriate array padding \cite{Hong_2016} reduces such conflict misses at the price of an increase in off-chip array size.

\paragraph{Exhibiting burst accesses by re-scheduling memory accesses} It may be possible to exhibit burst accesses by changing the access order of a given set of data addresses. If communications and computations are not separate \cite{Bayliss_2012}, then the execution's schedule is also the memory access schedule; a loop transformation that maximizes DRAM row and burst use is found. Such a process changing the loop's execution order, it may break temporal locality. When communications and computations are executed separately and on-chip memories are used as scratchpads \cite{Pouchet_2013}, the global memory access pattern does not need to match the on-chip access pattern. In this case, it is possible to improve memory bandwidth usage while keeping the same on-chip performance. 

\begin{figure}
	\centering
	\includegraphics[width=\columnwidth]{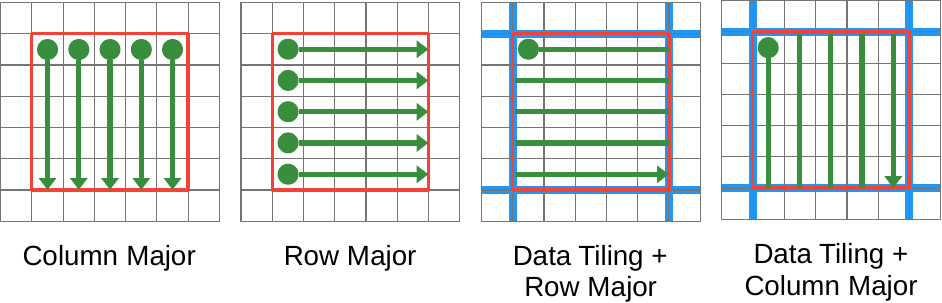}
	\caption{Memory layouts (non-tiled and tiled) for a 2-dimensional data space, and access patterns for a tile. Each dot is the start of a burst access, that spans until the arrow tip.}
	\label{fig:data-tiling-explanation}
\end{figure}

\paragraph{Re-allocating data in a burst-friendly layout} Regardless of the data layout, it is awlays possible to make burst accesses to memory, however the proportion of useful data accessed by such bursts may be low. It is possible to change the data layout to exhibit a high-usefulness burst access pattern. Data tiling \cite{Lam_1991} is one such technique; in the specific case of dense matrix product (GEMM), block matrix layout or \emph{data tiling} as illustrated on Figure \ref{fig:data-tiling-explanation} (c) is known to have excellent spatial locality \cite{Herrero_2006}. The dependence pattern of GEMM is such that the footprint of a tile can exactly fit a data tile, making it ideal layout for such an application. 
There is a trade-off between size and shape of data tiles and the usefulness of the burst accesses they permit, explored in works such as \cite{Ozturk_2009}.

\subsubsection{Increasing effective bandwidth by compressing data}
\keymsg{Compressing data increases the effective bandwidth, possibly at the cost of precision loss due to quantization.}

Another class of solutions around the bandwidth issue is using compression, whether lossy or lossless. Ozturk et al. \cite{Ozturk_2009} created a dynamic lossless compression engine that acts like a cache, where local data is compressed before being sent out to memory, and decompressed when coming from memory. Compression combines two advantages: it saves both memory and bandwidth, at the cost of using extra cores or on-chip area to perform it. A major pitfall of compression combined with data tiling is that it requires to read or write a full tile even to access a single point from it.

Lossy compression enhances throughput as well, at the price of errors. Maier et al.'s perforation \cite{Maier_2018} is a form of lossy compression, as is the well-known JPEG algorithm. Nakahara et al. \cite{Nakahara_2020} use it on CNN inputs; their method is to compress the CNN input on the host using the lossy JPEG algorithm, decompress it on FPGA chip, and perform the inference there. As expected, accuracy goes down as the JPEG quality factor decreases, and there is a trade-off with the obtained speedup.

Sun et al. \cite{Sun_2022}, to appear in 2022, tackle the same bandwidth problem as ours by using a mix of compression and data layout. Although this approach does not rely on polyhedral dependency analysis, it features the same base idea: group together data that is being used together.

\subsection{Automatic synthesis of optimized hardware}
\keymsg{Recent work in optimization of high-level synthesis has focused on memory allocation. Our work follows this trend.}

There is a longtime sustained community interest in automating optimization of hardware design, including high-level polyhedral optimizations (as opposed to low-level RTL tuning). Compilers have been developed specifically to this aim.

Early work on polyhedral compilation was already targeting hardware design; Le Verge, Mauras and Quinton in 1991 \cite{Le_Verge_1991} were using the Alpha language to automatically derive a systolic VHDL circuit from an Alpha polyhedral specification.

More recent work on HLS tools made it possible to explore tiling options for FPGA or ASIC accelerators (Pouchet et al., 2013 \cite{Pouchet_2013}) on various performance metrics such as latency, area, power consumption, memory bandwidth. Polyhedral transformations are done using specific tools, such as Pluto (Bondhugula et al., 2008 \cite{Bondhugula_2008}) that automatically identifies tilable loop nests and applies tiling.

The most recent advances in tools bear a focus on data movement and memory issues. The SODA framework \cite{Chi_2018} automatically generates a dataflow-like pipeline structure with FIFO-ordered off-chip accesses; the data is automatically transformed have a specific allocation for this to work. This approach turns the data layout into a specific pre-determined layout, independent of the actual dependence pattern, whereas ours finds a data layout and an access pattern for each accelerator in function of the dependence pattern.

Xiang et al., 2022 \cite{Xiang_2022} propose an approach that is complementary to ours, with an HLS code generator that infers the entire data movement and memory allocation, both off-chip and on-chip. This approach does not rely on fine-grain dependency analysis, and therefore has to keep the inner layout of each array; however, the location of each array in memory is carefully chosen so as to minimize the latency and maximize the bandwidth. Our approach is complementary: we make use of fine-grain dependency analysis to transform the inside of the arrays.

The next two sections explain in more detail the rationale and construction of our data layout, and the transformations a piece of code goes through to have it.

\section{Our method: Canonical Facet Allocation}
\label{sec:cfa}
This section details the core of our work, which principle is the following: by having different on-chip and off-chip data layouts, it is possible to have both on-chip parallelism (permitted by an adapted on-chip data layout) and low off-chip access latency (using a burst-friendly off-chip layout).

The next subsections explain the ideas that motivate our technique, and details how the off-chip data layout is built.

\subsection{Trade-off between read and write contiguity}
\label{sec:goals}

\begin{figure*}
	\begin{subfigure}{.2\textwidth}
		\centering
		\includegraphics{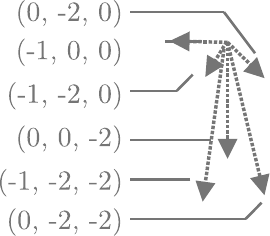}
		\caption{Dependence pattern (consumer-to-producer)}
		\label{fig:flow-in-deppattern}
	\end{subfigure}
	\begin{subfigure}{.4\textwidth}
		\centering
		\includegraphics{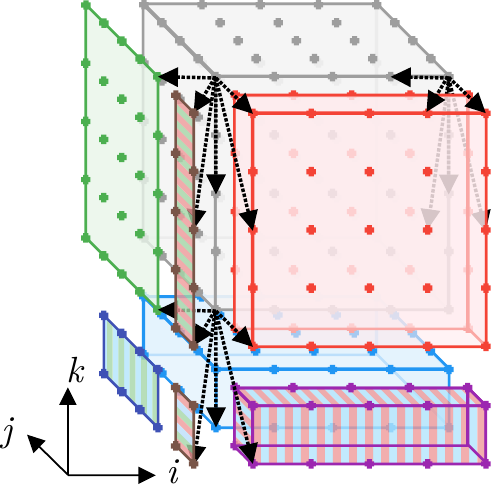}
		\caption{Iteration-wise flow-in set, split per producer tile. Consumer tile is in the background}
		\label{fig:flow-in}
	\end{subfigure}
	\begin{subfigure}{.4\textwidth}
		\centering
		\includegraphics{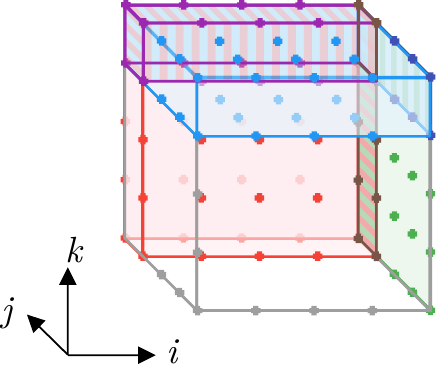}
		\caption{Iteration-wise flow-out set (facets)}
		\label{fig:flow-out}
	\end{subfigure}
	\caption{An instance of flow-in and flow-out sets. The flow-out set is the union of thicker versions of the tile faces (facets), while the flow-in set is composed of an union of either whole or partial facets. Some flow-in sets are adjacent in the iteration space; CFA reflects this adjacency in memory.}
	\label{fig:flow-inout}
\end{figure*}

It is not possible to have a data layout that would guarantee at the same time and for all cases, that both the flow-in and flow-out of a tile can be accessed in a single transaction. This is due to the fact that, in the general case, results produced by one tile are consumed by more than one tile, as on Figure \ref{fig:flow-out}, and each tile consumes results from multiple tiles as on Figure \ref{fig:flow-in}. 

Being able to read the whole flow-in data of a tile in a single burst transaction would require that a single contiguous region would be written to by multiple tiles. On Figure \ref{fig:flow-in}, a contiguous flow-in region would be written to from seven distinct iteration tiles. 

The same reasoning applies for the flow-out: if all the flow-out data from each tile was written to a single contiguous region of memory, the consumer tiles would pick their flow-in data from multiple regions and therefore make at least one access per producer tile. On Figure \ref{fig:flow-out}, each colored region belongs to the flow-in of different tiles.

The two above scenarios are extreme cases, where either the flow-in of or flow-out of a tile is a single contiguous region (using a single burst), but the other way is scattered. There is therefore a trade-off between write contiguity and read contiguity. Canonical Facet Allocation takes the following stance on this trade-off:
\begin{itemize}
  \item All write accesses are burst accesses (there are no element-wise writes), 
  \item Minimize the number of read transactions, whether they are element-wise or bursts.
\end{itemize}

The minimization objective provides no guarantee of absence of element-wise transactions, which sometimes may be unavoidable.

\subsection{Separating off-chip and on-chip data layouts}

It is, in general, not necessary that the off-chip data layout be the same as the on-chip layout. Indeed, both have a different objective: the on-chip layout should maximize access parallelism with low port contention, while the off-chip layout should allow minimal access latency and maximum throughput. We therefore propose to use two distinct layouts.

On-chip allocation has been the subject of many prior work, whether they are optimizing techniques or their automation; we assume it is already possible to find a suitable on-chip allocation to maximize parallel accesses. The core of our contribution is on off-chip allocation: we propose an off-chip allocation scheme adapted to tiled loop nests.

\subsection{Principles of construction of off-chip allocation}

We propose an allocation scheme that leverages multiple levels of contiguity. Relying on the shape of iteration space tiles, our allocation scheme offers contiguity for data produced by a tile (full-tile contiguity and intra-tile contiguity) as well as between adjacent tiles (inter-tile contiguity).

The data layout we propose honors these contiguity properties at the same time. The next subsections explain how we build it thanks to the combination of three techniques:
\begin{itemize}
  \item multi-projection of each tile,
  \item data tiling,
  \item array dimension permutation.
\end{itemize}

\subsection{Definitions}

We use the following terminology in the next subsections:

\textbf{Projection}: We use a restricted definition of projection, considering only orthogonal projections. They are used to strip one dimension from a multi-dimensional space. For instance, the projection $p_k$ such that $$p_k(i_0, i_1, \dots, i_k, \dots, i_d) = (i_0, i_1, \dots, i_{k-1}, i_{k+1}, \dots, i_d)$$ removes the $k$-th dimension from an $d$-dimensional space.

\textbf{Tile}: A tile is an hyperrectangle, subset of the iteration space. Its size is $t_1 \times \dots t_d$ in general; in the 3-dimensional example of Figure \ref{fig:flow-inout}, tile size is respectively $t_i$, $t_j$, $t_k$ along the $i$, $j$, $k$ axis. A tile has coordinates $i_1, \dots, i_d$ along each axis; in the example, these are $ii$, $jj$ and $kk$.

\textbf{First-level neighbor}: A first-level neighbor of a tile is a neighbor that is reached with a move along a single canonical axis. For instance, $(ii, jj + 1, kk)$ is a first-level neighbor of $(ii, jj, kk)$, but $(ii + 1, jj + 1, kk)$ is not.

\textbf{Second-level neighbor}: A second-level neighbor of a tile is a neighbor such that is reached with a move along exactly two distinct canonical axes. For instance, $(ii - 1, jj + 1, kk)$ is a second-level neighbor of $(ii, jj, kk)$, but $(ii + 1, jj + 1, kk + 1)$ and $(ii + 1, jj, kk)$ are not.

\textbf{$k$-th level neighbor}: By extension of the above two definitions, a $k$-th level neighbor of a tile is a neighboring tile that is reached with a move along exactly $k$ canonical axes.

We will use the following notations:
\begin{itemize}
\item $E\subset\mathrm{vect}\left(\vec{e_{1}},\dots,\vec{e_{d}}\right)$
: Iteration space of $d$ dimensions
\[
E=\left\{ \vec{x}=\left(x_{1},\dots,x_{d}\right):0\leqslant x_{1}<N_{1},\dots,0\leqslant x_{d}<N_{d}\right\} 
\]
\item $\vec{B_1}, \dots, \vec{B_p}$ : dependence vectors, such that rectangular tiling is legal. We assume all dependence vectors are backwards in all dimensions: $\forall i, j: \vec{B_i}\cdot\vec{e_j} \leqslant 0$.
\item $N_{1} \times \dots \times N_{d}$ : iteration space size
\item $t_{1},\dots,t_{d}\in\mathbf{N}^*$ : tile sizes
\end{itemize}

\subsection{Hypotheses}

We make the following hypotheses:

\textbf{Uniform dependencies}: It is, in general, impossible to predict the shape of the flow-in set of a tile when dependencies are not uniform. With uniform dependencies, the flow-in set of all tiles has the same shape, and the same goes for the flow-out set. Many programs under active research fall under this hypothesis, ranging from stencils to convolutions.

Also, this hypothesis does not imply that all memory accesses are uniform - this only applies to accesses to read-write arrays (those that hold intermediate results).

\textbf{Rectangular tiling}: We assume tiles are rectangular in all dimensions; using polyhedral tools, it is notably possible to change the iteration space basis so that rectangular tiling becomes legal. Therefore, before applying CFA, we expect such a pre-processing to have been done if necessary. This does not imply a lack of generality.

\textbf{Non-sparse data}: It does not make sense to apply CFA on highly sparse data. The data layout of CFA is dense per construction, so that uniform dependencies yield uniform memory accesses. Using it with sparse data would lead to a significant amount of avoidable redundant data transfers. Sparse data should use a sparse representation instead.

\subsection{Contiguity along multiple spatial directions: multi-projection}

The first aspect of Canonical Facet Allocation is to make multiple projections of the iteration space. This makes it possible to have contiguity in single-dimensional memory along multiple spatial directions. The next two paragraphs explain why and how CFA achieves this.

\subsubsection{Rationale}
To get the best spatial locality, data should be laid out in memory the same way as the producer iterations are in the iteration space. Given that memory is abstracted to a one-dimensional space, in which it is not possible to have a multi-dimensional layout in memory. However, it is also true that:
\begin{itemize}
    \item Not all data produced within an iteration tile needs to go through global memory - only flow-in and flow-out data need to,
    \item Only select neighbors of those iterations in flow-in / flow-out sets also belong to these sets. 
\end{itemize}

An ideal memory allocation in terms of contiguity would be similar to a mathematical continuous function: two neighboring flow-in / flow-out iterations in \emph{any} direction would also be neighbors in memory. Since the memory is one-dimensional, there is no such allocation. There is at most one direction per data space where this property can be verified, which is its direction of contiguity. It is however possible to use multiple data spaces, each one with a different direction of contiguity.

Because each flow-in / flow-out data point only needs specific directions of contiguity to exhibit spatial locality, we carefully choose the data that belongs to each data space. Visually, on Figure \ref{fig:flow-out}, one would like to cut the flow-in / out data into three pieces, one per face of the cube that has flow-out data. This is what multi-projection achieves.

Multi-projection consists in creating multiple data spaces from a single iteration space, each data space corresponding to some projection of the iteration space. In the example of Figure \ref{fig:flow-out}, there is a data space for each of the canonical hyperplanes $(i, j)$, $(i, k)$ and $(j, k)$. 

The next two paragraphs state how each projected data space is created, first with an example, then in the general case.

The idea is to make data spaces that are as thick as the dependence pattern "plunges" into the neighboring tiles.

\subsubsection{Construction example}

We start with a visual way to construct the data spaces, from a 3-dimensional iteration space, from Figure \ref{fig:flow-inout}. The idea to construct the projected data space is, for every canonical hyperplane:
\begin{enumerate}
\item To determine how thick the facet (and consequently the data space) should be, and 
\item To create a function that maps iteration coordinates to data space coordinates.
\end{enumerate}

On Figure \ref{fig:flow-out}, the part of the flow-out parallel to hyperplane $(i, j)$, in light blue, has a thickness of 2; otherwise said, consumer tiles will need the result of the two uppermost $(i, j)$ planes ($k \in \lbrace 3,4 \rbrace$). The dual way to see it is the flow-in (Figure \ref{fig:flow-in}): when moving the dependence pattern along the bottom plane of a consumer tile, the iterations this plane depends on (part of which the blue slab below the consumer tile) are located two planes below it.

The thickness of each data space is calculated using the dependence pattern: it is the maximum length of every dependence vector along the normal vector to the hyperplane we are projecting on. In the example of Figure \ref{fig:flow-inout}, the dependence pattern is on Figure \ref{fig:flow-in-deppattern}. For hyperplane $(j, k)$, it is the maximum absolute value of the component along the $i$ axis of every dependence vector. We determine it to be 1: only the rightmost $(j, k)$ plane of iterations is needed. Therefore, the data space created from the $(j, k)$ hyperplane will be a simple two-dimensional array, that may be declared as \texttt{Dtype facet\_i[$N_j$][$N_k$]}. We name it "facet" because it will ultimately contain data tiles holding entire facets.

To give coordinates to the iterations' results in the data space, we use a simple projection of the rightmost face of a tile: $$p_i(i, j, k) = (j, k)$$ which domain is that rightmost face of each tile. As the tile size is 5, we get: $$D(p_i) = \left\lbrace (i, j, k) : i \equiv 4 \mod 5\right\rbrace$$

Note that, although we do not do it at this point, the above projection function can be translated to code. Generating code of this function in its domain would result in a \texttt{for} loop that browses the rightmost face of each tile (in this example, tile  size is $5$ in each dimension): \texttt{for($j=5jj$ to $5jj+4$)\{ for($k=5kk$ to $5kk+4$)\{ facet\_i[$j$][$k$] = iteration\_result[4][$j$][$k$]\}\}}. 

For hyperplane $(i, j)$, the maximum absolute value of the component along the $k$ axis of every dependence vector is 2. Therefore, the data space to be created will consist of the two uppermost $(i, j)$ planes of every tile.

To create the mapping between iteration and data spaces, we use a modified projection, called a modulo projection: instead of getting rid of the component along the $k$ axis, this projection replaces the $k$ axis by $k \mod 2$: $$p_k(i, j, k) = (i, j, k \!\!\!\mod 2)$$
The domain is the two uppermost planes: $$D(p_k) = \left\lbrace (i, j, k) : 3 \leqslant k \!\!\!\mod 5 \leqslant 4 \right\rbrace$$

\subsubsection{General case}

In the general case, i.e. for $d$-dimensional spaces, we determine the thickness of each facet (i.e. each data space) the same way as above. The thickness of each facet is determined by the longest dependence vector along the direction normal to that face. Assuming dependence vectors $\vec{B_{1}}, \dots, \vec{B_{p}}$, the thickness of facet normal to $\vec{e_{k}}$ is given by:
$$w_{k}=\max_{q\in \intset{1,p} }\left|\vec{e_{k}} \cdot \vec{B_{q}}\right|$$

A proof that this is sufficient to hold all flow-in/flow-out data is located in annex.

We then take canonical \emph{modulo} projections to map the iterations to the data space:
$$ p_k(x_1,\dots,x_d) = (x_1, \dots, x_{k-1}, x_k \!\!\!\mod w_k, x_{k+1}, \dots, x_d) $$
that are defined over the $w_k$ last planes of a tile:
$$ D(p_k) = \left\lbrace (x_1, \dots, x_d) : t_k - w_k \leqslant x_k \!\!\!\mod t_k \leqslant t_k - 1 \right\rbrace $$

\subsubsection{Single-assignment scheme}

The projections introduced in this section cause facets from tiles along a canonical axis to overwrite each other's data (all tiles along a canonical axis share the same memory locations for the data from the hyperplane normal to that axis).
For instance, with the example of Figure \ref{fig:flow-out}, all tiles along the $i$ axis share the same projection of the $(j, k)$ plane in the \texttt{facet\_i} array.

The consequence of this overwriting is the possible introduction of memory dependences that may render the schedule of tiles illegal (data from a tile could be overwritten while it is still needed). In order not to break the schedule's legality, we choose to replicate the data spaces so as to make each data space a \emph{single-assignment} space. This means each iteration tile will write in a memory space distinct from that of any other tiles. 

Single-assignment spaces are created by introducing an extra dimension in the facet arrays that corresponds to the direction normal to the projection hyperplane. Continuing with Figure \ref{fig:flow-out} example, the facet corresponding to $(j, k)$ hyperplane is augmented with a dimension on the $i$ axis, with so many entries as there are tiles on the $i$ axis. The corresponding array becomes \texttt{facet\_i[$N_i$/5][$N_j$][$N_k$]}.

\subsection{Flow-in from first-level neighbors: Full-tile contiguity}

We want to ensure that each facet from each tile is written in the form of a burst access, which we call \emph{full-tile contiguity}. Data tiling gives us this level of contiguity. The idea is to mirror the iteration space tiles on the data space; data tiling then allows to write each projected data space (facet) in a single burst access. 

This transformation will improve both write and read performance: Flow-out facets are almost entirely used by the tile they are immediately adjacent to.

\begin{figure}
	\centering
	\includegraphics[width=0.5\columnwidth]{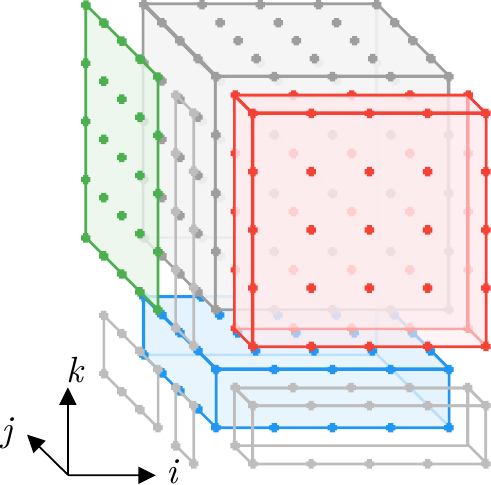}
	\caption{Flow-in facets from first-level neighbors. Each facet is a data tile, i.e. a contiguous block of memory.}
	\label{fig:facets}
\end{figure}

The next two subsections give an example and the general way to apply data tiling in the projected data spaces.

\subsubsection{Example}
We can continue with the example of Figure \ref{fig:flow-inout}. In this example, we have built data spaces for each facet:
\begin{itemize}
    \item \texttt{facet\_i[$N_i$/5][$N_j$][$N_k$]}
    \item \texttt{facet\_j[$N_j$/5][$N_i$][$N_k$][2]}
    \item \texttt{facet\_k[$N_k$/5][$N_i$][$N_j$][2]}
\end{itemize}

We would like that the data from each facet of each tile to be written with a single burst access. For hyperplane $(j, k)$, which is written to \texttt{facet\_i}, we would like the destination space of projection $p_i$ to be contiguous in memory, so that the code generated by the projection makes a single burst. We call this \emph{full-tile contiguity}.

The destination space (image) of $p_i$ is: \\ $M_i=\left\lbrace (i, j) : 5ii \leqslant i < 5(ii+1) \wedge 5jj \leqslant j < 5(jj+1) \right\rbrace$, \\ which corresponds to a flattened version of the iteration tile $(ii, jj)$ onto the data space. The rectangular subset of the array \texttt{facet\_i} that corresponds to $M_i$ needs to be contiguous in memory.

The array \texttt{facet\_i} will therefore be split into tiles of size $5\times 5$, resulting in the following array:
\begin{center} \texttt{facet\_i[$N_i$/5][$N_j$/5][$N_k$/5][5][5]} \end{center}

\subsubsection{General case}

To mirror iteration tiles on the data space, we apply data tiling with the same tile sizes as the iteration space: the projection normal to $\vec{e_k}$ is therefore tiled with dimensions $(t_1, \dots, t_{k-1}, t_{k+1}, \dots, t_d)$.

Tiling the $i$-th dimension of an array with size $t$ consists in replacing the $i$-th dimension by two dimensions, namely the quotient and the remainder of the Euclidean division of the original dimension by $t$. This operation gives $2(d-1)$-dimensional data spaces from a $(d-1)$-dimensional projection of the iteration space. All the quotient dimensions are moved first, and the remainder dimensions are moved last. We call the quotient dimensions the \textbf{\emph{outer} dimensions}, and the remainder dimensions the \textbf{\emph{inner} dimensions}.

Tiling a 3-dimensional data space $A$ with tile sizes $(t_i, t_j, t_k)$ would for instance be done with the following mapping from the original 3-dimensional to the tiled 6-dimensional array:

$$ 
\begin{array}{l}
A[i][j][k] \mapsto 
A'
\displaystyle{
\left[\frac{i}{t_i} \right]
\left[\frac{j}{t_j} \right]
\left[\frac{k}{t_k} \right]
[i \texttt{\%} t_i]
[j \texttt{\%} t_j]
[k \texttt{\%} t_k]}
\end{array}$$

This transformation is composed with the projection function of the previous section to give the actual allocation.

Data tiling provides a contiguity guarantee if the entire facet is read. As on Figure \ref{fig:facet-extensions}, we need to read only a subset of a facet from a second-level or third-level neighbor. Finer tweaking of the data layout is necessary so that these subsets are contiguous. The next subsections deal with these cases and brings in additional levels of contiguity.

\subsection{Flow-in from second-level neighbors: Inter-tile contiguity}

CFA mirrors contiguity across iteration tiles into contiguity across data tiles. This is called "facet extensions", as a burst access reading from data tile (holding a facet's data) can be extended to fetch a neighboring tile's data.

\begin{figure}
	\centering
	\includegraphics[width=0.5\columnwidth]{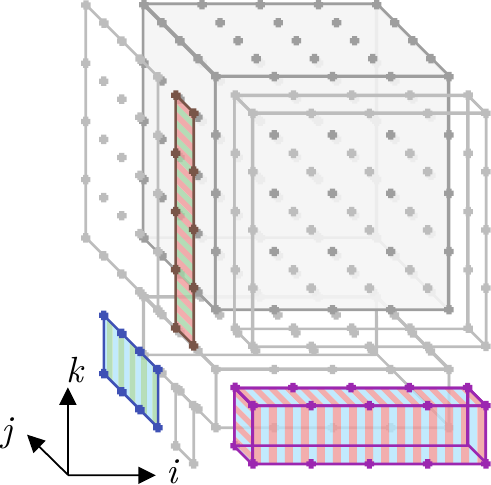}
	\caption{Flow-in iterations from second-level neighbors of an iteration tile. CFA places these in memory next to a data tile already read to access them as ``extensions''.}
	\label{fig:facet-extensions}
\end{figure}

In the general case, part of the flow-in iterations of a tile are located in its second-level neighbors, such as those as shown in Figure \ref{fig:facet-extensions}. 

\begin{figure}
	\centering
	\includegraphics[width=\columnwidth]{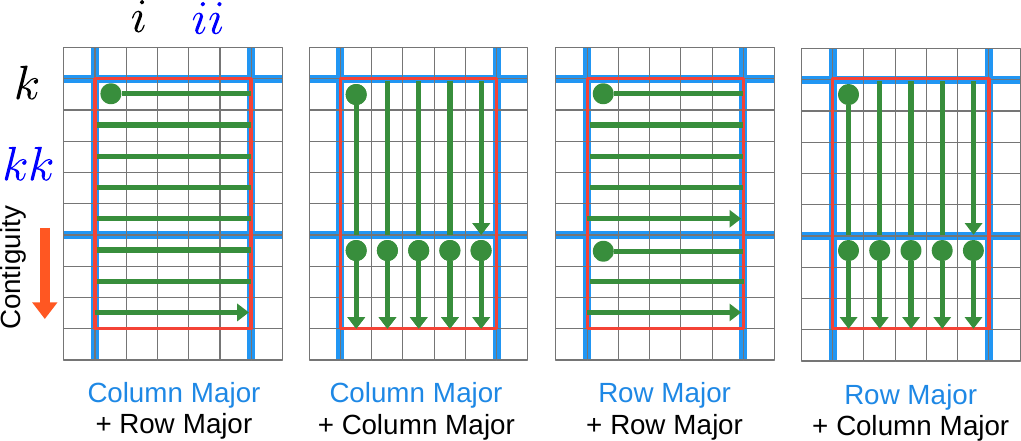}
	\caption{The four possible data layouts for a 2-dimensional array in tiled data layout (inter-tile + full-tile layouts), and access pattern of a footprint that spans across data tiles. We seek inter-tile contiguity along direction $k$, which only column + row-major layout allows.}
	\label{fig:cross-tile-contiguity}
\end{figure}

We applied multi-projection and data tiling so that two neighboring tiles in memory are necessarily first-level neighbors in the iteration space. Contiguous reads from a first-level to a second-level iteration neighbor are possible, since a second-level neighbor is a first-level neighbor of a first-level neighbor. Such contiguity enables a cross-tile border read in a single burst access, as is shown for a two-dimensional array on Figure \ref{fig:cross-tile-contiguity}. We call this level of contiguity \emph{inter-tile contiguity}.

This can be obtained by swapping the dimensions of the facet arrays. The next paragraphs explain how.

\subsubsection{Example}
\label{sec:ex-inter-tile}

We take back the example of Figure \ref{fig:flow-inout}, of which the part of the flow-in data in second-level neighbors is shown in Figure \ref{fig:facet-extensions}. The producer iterations are part of two facets at the same time. We therefore have to:
\begin{enumerate}
    \item Choose a direction of contiguity for each facet, and
    \item Select the right facet to read each extension from.
\end{enumerate}

If we call $E_{(ii, jj - 1, kk - 1)}$ the set coming from tile $(ii, jj - 1, kk - 1)$ (purple slab on Figure \ref{fig:facet-extensions}), we notice that this slab is a subset of both the $(i, j)$ and $(i, k)$ hyperplanes, so we can read it from both \texttt{facet\_j} and \texttt{facet\_k} arrays. 

We choose that the direction of contiguity for the data space projected from the $(i, k)$ hyperplane to be the $k$ axis, and to read the extension $E_{(ii, jj - 1, kk - 1)}$ from \texttt{facet\_j}. Figure \ref{fig:cross-tile-contiguity} shows the four possible layouts for the \texttt{facet\_j}: only column-major as a data tile layout and row-major as an intra-tile layout will allow reading $E_{(ii, jj - 1, kk - 1)}$ as a contiguous extension of the $(i, k)$ facet of tile $(ii, jj - 1, k)$. We choose this layout, and swap the dimensions of the \texttt{facet\_j} array to match it. Column-major inter-tile layout means that dimensions are ordered this way: \texttt{kk}, \texttt{ii}; and row-major intra-tile layout gives the \texttt{i}, \texttt{k} order.

Accessing the \texttt{facet\_j} array is therefore done with \texttt{facet\_j[jj][ii][kk][k][i][2]}.

The result is that the purple and red slabs of Figure \ref{fig:flow-in} can be read in a single, merged burst, from \texttt{facet\_j}: they are contiguous along the $k$ axis.

The same process can be repeated with the other facets.

\subsubsection{General case}

A tiled data space, as on Figure \ref{fig:cross-tile-contiguity}, has two dimensions corresponding to each canonical axis (one for tile coordinates, and one for intra-tile coordinates). By convention, if $i$ is an axis, let's designate \texttt{ii} the tile coordinate along that axis, and \texttt{i} the coordinate of a cell along the $i$ axis inside a tile (this convention is followed on Figure \ref{fig:cross-tile-contiguity}). 

The direction of inter-tile contiguity for a facet has to be chosen among those axes that are projected.

For a given projection, to make tiles along the $i$ axis contiguous, then the \texttt{ii} dimension is moved as the last of the outer dimensions to be enumerated. Enabling contiguous reads is granted by promoting the \texttt{i} dimension the first of the inner dimensions to be enumerated.

Once the direction of inter-tile contiguity is picked, those parts of the flow-in sets that have been made contiguous can be merged to be read in a single burst.

\subsection{Flow-in from third-level neighbors: Intra-tile contiguity}

\begin{figure}
	\centering
	\includegraphics[width=0.5\columnwidth]{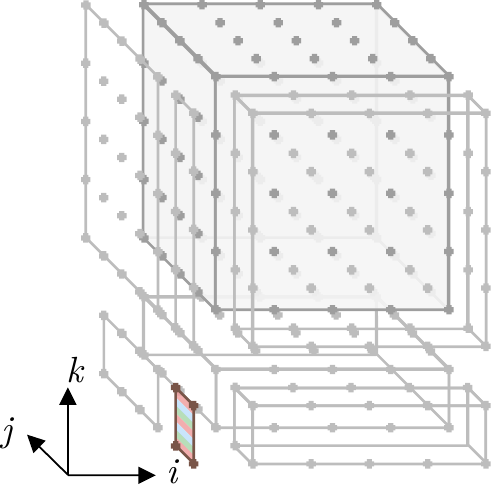}
	\caption{Flow-in from third-level neighbor (pattern of Figure \ref{fig:flow-inout}. CFA places these four points contiguously in memory within a data tile (intra-tile contiguity).}
	\label{fig:singleton}
\end{figure}

The subsets of flow-in coming from third-level neighbors, as is the set on Figure \ref{fig:singleton}, are in general not contiguous in memory to data already accessed from first- or second-level neighbors. Still, this set may be read as a single contiguous burst. This section deals with the specific case of a three-dimensional iteration space.

In a 3-dimensional iteration space, there is only one flow-in set from a single third-level neighbor. It has a constant number of points, only a function of the dependence pattern. We call $S_3$ this subset.

For instance, on Figure \ref{fig:singleton}, $S_3$ is the subset of iterations coming from tile $(ii - 1, jj - 1, kk - 1)$ with $i=4$, $j \in \lbrace 3,4 \rbrace$ and $k \in \lbrace 3,4 \rbrace$. This subset needs to be contiguous within one of the three projected facets.

It is possible to make $S_3$'s data contiguous in memory by changing the order of the inner dimensions of \texttt{facet\_k} (projection of $(i, j)$ planes). Since this facet only contains iterations with $k \in \lbrace 3,4 \rbrace$, we make $S_3$ contiguous using the order of dimensions $kk, jj, ii, i, j, k$. Then, the subset of Figure \ref{fig:singleton} is contiguous within \texttt{facet\_k}: for any $i$, the points $(i, 3, 3)$, $(i, 3, 4)$, $(i, 4, 3)$ and $(i, 4, 4)$ are consecutive in memory. We can thus fetch them in a single burst.

The final layout of our arrays is therefore:

\begin{itemize}
    \item \texttt{facet\_i[ii][jj][kk][j][k]}
    \item \texttt{facet\_j[jj][ii][kk][k][i][j \% 2]}
    \item \texttt{facet\_k[kk][jj][ii][i][j][k \% 2]}
\end{itemize}

\subsection{Case of $k$-th level neighbors}

Although full, inter and intra-tile contiguity are always possible for first, second and third-level neighbors in a 3-dimensional space, it may not be in a 4 or higher-dimensional iteration space: the number of $k$-th level neighbors of a tile is $C_k^d$, which is higher than $d$, the maximum number of projections (i.e. of directions of contiguity).

\section{Code generation}

As part of our work, we have written a proof-of-concept source-to-source compiler pass that takes a C program corresponding to a tile and transforms it so that it uses CFA for global memory accesses.

Transforming the program to use CFA involves two main steps: analyzing the program's dependencies and calculating the facets, then transforming the program so that it uses the facets. Our proof-of-concept compiler pass sits within a compiler framework that targets high-level synthesis engines. Therefore, the code produced by our pass is turned into an FPGA accelerator using a synthesis tool such as Vitis HLS or Catapult.

This section explains how the program is transformed to include off-chip accesses to facets.

\subsection{Overview: compiler pass}

The flow of our compiler pass is as follows: it takes the the polyhedral representation of a program as input, under the hypotheses stated in the previous section. It determines what the facets are, under the form of sets of points. It then generates loops to scan these sets of points contiguously (copy-in / copy-out code), and wraps the tile's code with this copy-in and copy-out code. The copy-in/out code accesses global memory in CFA layout and turns it into the original program's layout for fast on-chip access.

\begin{figure}
    \centering
    \includegraphics[width=\columnwidth]{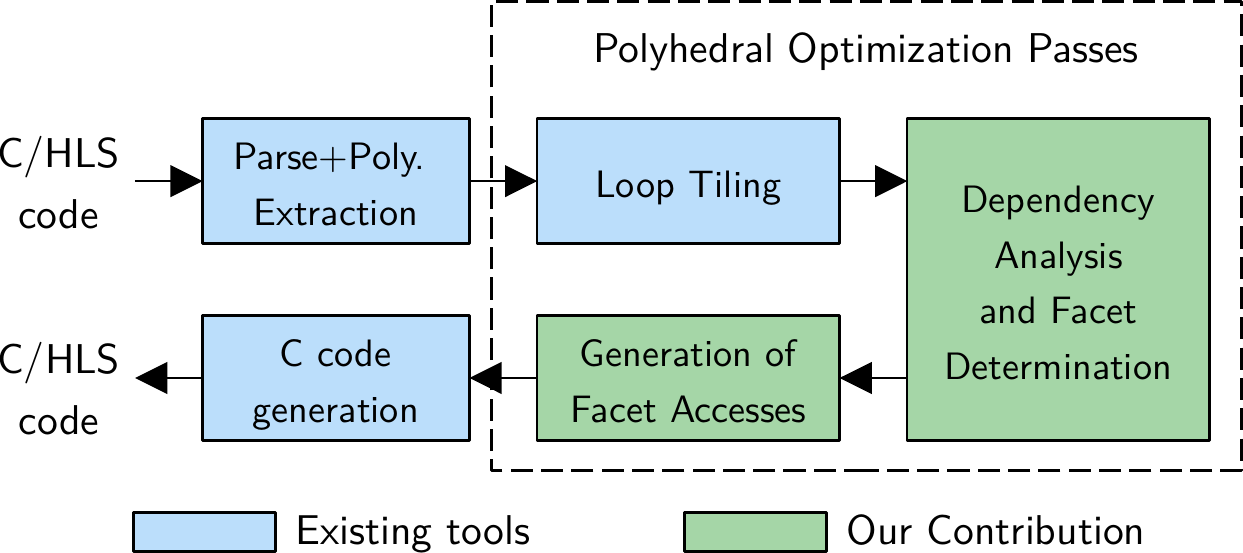}
    \caption{Polyhedral compiler flow including the CFA pass (shown broken into two).}
    \label{fig:compiler_flow}
\end{figure}

The following subsections explain the transformations that are applied to the code and how copy-in/out code is generated.

\subsection{Determining the facets and their layout}

The technique described in Section \ref{sec:cfa} is applied in order to determine the facets and their layout. Existing tools are used to do this: tiling is done using Pluto \cite{Bondhugula_2008}; the Integer Set Library (ISL) is used to represent the iteration and data spaces, as well as the affine relations between them.

\subsection{Copy-in/out code generation}

CFA itself is only an allocation scheme for the data, which means it states \emph{where} each datum is placed, not in \emph{which order} the data is read or written. This subsection explains how we can make an access pattern that benefits from the contiguity properties of CFA, by issuing long burst accesses.

\subsubsection{Making a contiguous flow-in/flow-out access pattern}

We need to generate an access pattern that will take out the flow-in data from facets, and conversely, write the flow-out data into facets. This access pattern is supposed to be as contiguous as possible, and we expect the longest possible burst accesses.

Flow-out facets are entirely written with one transaction per facet - all the flow-out data is contained.

\begin{figure}
\centering
\includegraphics[width=\columnwidth]{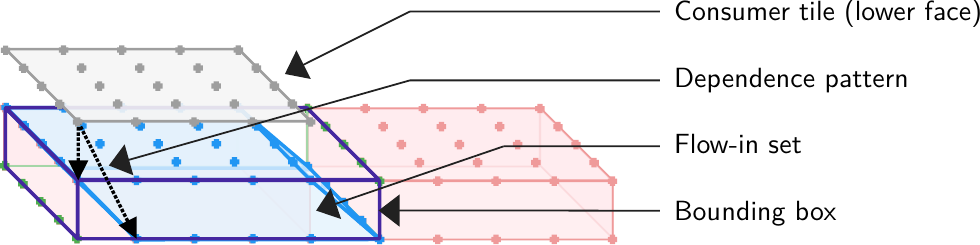}
\caption{Taking a rectangular over-approximation of the actual flow-in set incurs redundant reads. Contiguous data (facets) are in red: an over-approximation may be necessary to preserve access contiguity.}
\label{fig:TobleroneEffect}
\end{figure}

The intersection of the actual flow-in set with each facet may not be exactly a rectangle, and is perhaps not contiguous inside that facet. For this reason, a rectangular over-approximation of the set of accessed data is taken, like on Figure \ref{fig:TobleroneEffect}. That superset may span across facets from multiple iteration tiles - in this case, cross-tile contiguity ensures that a single transaction can bring all the data on-chip.

Given that a bounding box of a subset is contained in the bounding box of the containing set, doing so incurs less overhead than making a rectangular bounding box of the whole flow-in data.

\emph{Correctness:} For flow-out data, due to the choice of using a tile-wise single-assignment allocation (no two different tiles write in the same memory cells), this over-approximation poses no correctness issues. For flow-in accesses, we must ensure this over-approximation doesn't break correctness (conflicts caused by two iterations sharing the same on-chip cell, while one is not part of the flow-in data). Filtering out the unneeded data coming from the bounding box is therefore necessary by adding a guard to the copy-in code.

\subsubsection{Generating Burst-Capable Code}

The code to be generated to fetch and write flow-in/flow-out sets has to trigger burst accesses on its memory controller. We generate synthesizable high-level synthesis (HLS) code, where bursts are inferred from \texttt{for} loops. The following conditions are sufficient for a burst to be inferred:

\begin{itemize}
    \item The number of addresses accessed is explicit (e.g., the trip count of the copy loop nest is constant),
    \item If the target array size is known, the memory addresses are contained within its bounds,
    \item The addresses accessed are be consecutive (e.g., with a pointer increment),
    \item If applicable, the copy loop is pipelined with an initiation interval of 1 (assuming on-chip arrays are partitioned in such a way that this is possible).
\end{itemize}

Using the rectangular over-approximation of facet accesses, we can generate  rectangular loop nests, that are coalesced into a single loop to force the inference of a single burst instead of a series of shorter bursts. as on Figure \ref{fig:copyCode}.

The copy code features two address generators, for off-chip and on-chip data. Off-chip address calculation is straightforward: when accesses are contiguous, one simply needs to provide the HLS tool with a pointer that starts at the beginning of the memory region to be accessed, and increment it. On-chip address calculation is performed at each cycle.

The resulting code for each facet is as on Figure \ref{fig:copyCode}, and is sufficient to get burst accesses using a commercial HLS tool like Vitis HLS.

\begin{figure}
    \centering
    \begin{lstlisting}[style=CStyle]
copy:
float* offChipAddr = &facet_0[l][n][m][0] + 0;
for(int I = 0; I <= 9215; I = I + 1) {
	#pragma HLS PIPELINE II=1
	int c6 = I / 96
	int c7 = I
	*offChipAddr = A_local[(96*n + c7 + 0)
	offChipAddr = offChipAddr + 1;
}\end{lstlisting}
    \caption{Merger of two loops from which a burst access is inferred. on-chip addresses are random, while off-chip addresses are consecutive.}
    \label{fig:copyCode}
\end{figure}

\subsection{Generating HLS code}

The final step in CFA code generation is to generate a three-step coarse-grain pipeline:
\begin{itemize}
    \item The first step reads the flow-in data from global memory in CFA allocation, and turns it into local allocation,
    \item The second step is tile execution,
    \item The last step is writeback from the accelerator to global memory.
\end{itemize}

This is implemented under the form of a function, as in Figure \ref{fig:toplevel}. Note the DATAFLOW pragma, which is used by the HLS engine to generate a coarse-grain pipeline where the three functions \texttt{read}, \texttt{execute} and \texttt{write} are executed in parallel, each function for a different tile.

\begin{figure}
    \centering
    \begin{lstlisting}[style=CStyle]
void toplevel(int i, int j, int k, float* facetIJ, 
  float* facetIK, float* facetJK) {
#pragma HLS INTERFACE m_axi port=facetIJ
#pragma HLS INTERFACE m_axi port=facetIK
#pragma HLS INTERFACE m_axi port=facetJK
#pragma HLS DATAFLOW
    float buf1[TS*TS]; 
    float buf2[TS*TS]; 
    read(i, j, k, facetIJ, facetIK, facetJK, buf1);
    execute(i, j, k, buf1, buf2);
    write(i, j, k, buf2, facetIJ, facetIK, facetJK);
}\end{lstlisting}
    \caption{Top-level function, assuming tile size is $\mathtt{TS}$,
    local memories are $\mathtt{TS}^2$ big and the iteration space is
    $\mathtt{TS}^3$. This is the case of iterative stencils.}
    \label{fig:toplevel}
\end{figure}

\section{Evaluation}
\label{sec:eval}
This section evaluates Canonical Facet Allocation by answering three questions with respect to the state of the art:
\begin{itemize}
  \item Does Canonical Facet Allocation use all the available memory bandwidth?
  \item What is the improvement in effective bandwidth of CFA?
  \item Is there an area overhead due to using CFA versus single-layout allocations?
\end{itemize} 

\subsection{Experimental protocol}

Experiments were carried out on a variety of uniform-dependence benchmarks listed in Table \ref{tab:benchmarks}. Benchmarks such as iterative stencils  update an array in place, and differ by the dimensions of the iteration space and shape of the dependence pattern.

\begin{figure}
\centering
\includegraphics[width=0.7\columnwidth]{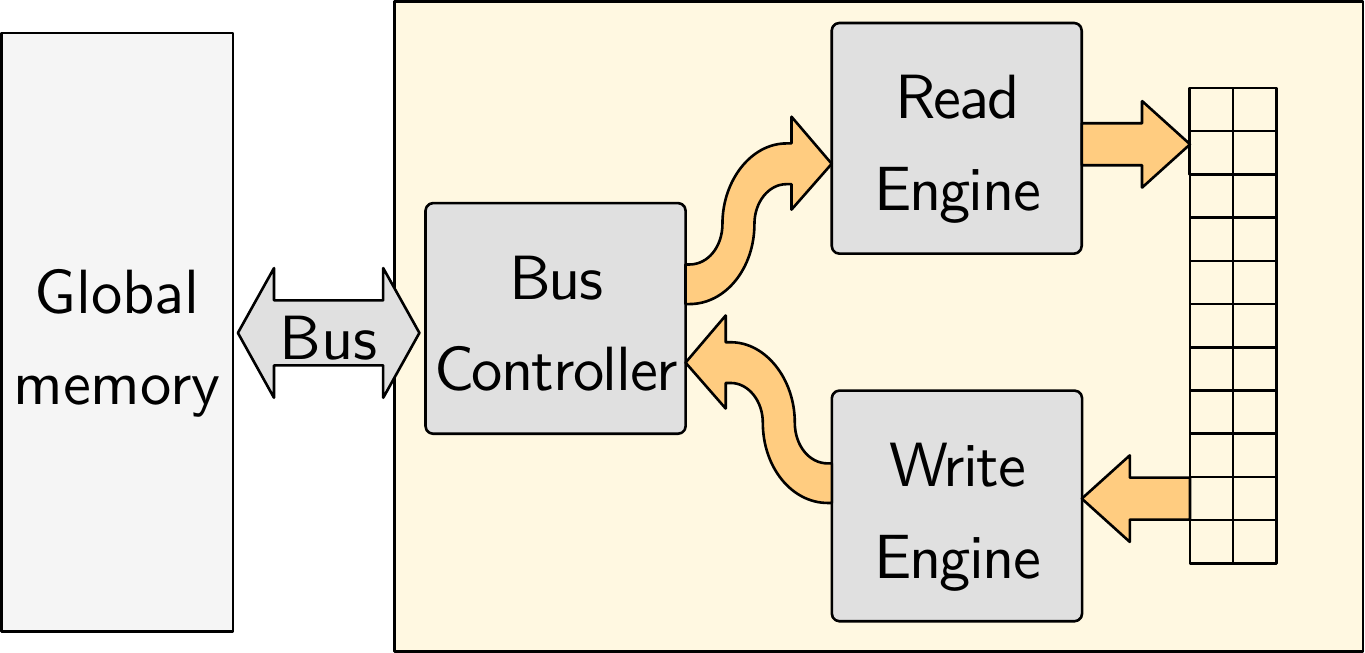}
\caption{Memory-bound accelerator used for benchmarking: only the read and write engines are implemented.}
\label{fig:CFABenchmarkHW}
\end{figure}

The platform used is a \emph{Xilinx Zynq ZC706} board, including an \emph{xc7z045ffg900-2} FPGA. The test accelerators only comprise read and write parts, the structure being that of Figure \ref{fig:CFABenchmarkHW}. Every baseline has been tested by connecting it to a single AXI high-performance port mapped to DRAM (port HP0); the frequency of every design is 100.00 MHz, the AXI bus is 64-bit wide, and the data type transferred over the bus is 64-bit double-precision IEEE floating point numbers.

\begin{table}
\centering

\begin{tabular}{|>{\centering\arraybackslash}m{.3\columnwidth}|>{\centering\arraybackslash}m{.1\columnwidth}|>{\centering\arraybackslash}m{.15\columnwidth}|>{\centering\arraybackslash}m{.25\columnwidth}|}
\hline
Dependence pattern & Nb of deps & Tile Sizes & Equivalent Application \\
\hline
\texttt{jacobi2d5p} & 5 & $16^3\rightarrow128^3$ & Laplace equation \\
\hline
\texttt{jacobi2d9p} & 9 & $16^3\rightarrow128^3$ & $3\times3$ convolution \\
\hline
\texttt{jacobi2d9p-gol} & 9 & $16^3\rightarrow128^3$ & 2nd-order finite difference\\
\hline
\texttt{gaussian} & 25 & $4\times16^2\rightarrow4\times128^2$ & $5\times5$ Gaussian Blur \\
\hline
\texttt{smith-waterman} \texttt{-3seq} & 7 & $16^3\rightarrow128^3$ & Alignment of 3 sequences \\
\hline

\end{tabular}
\caption{Benchmarks used for testing CFA. Equivalent applications have the same computational dependence pattern as the benchmark and would show similar performance.}
\label{tab:benchmarks}
\end{table}

\subsubsection{Baselines}

We considered the following baselines for comparison with CFA:
\begin{itemize}
  \item \textbf{Original Layout} (as done by Bayliss et al. \cite{Bayliss_2012}): a best-effort burst access pattern is determined under the original allocation. This access pattern does not issue any redundant reads, possibly at the expense of contiguity.
  \item \textbf{Bounding Box} (as done by Pouchet et al. \cite{Pouchet_2013}): a rectangular bounding box around the flow-in and flow-out data is taken so as to exhibit burst transfers; part of the bounding box is unused and redundantly transferred.
  \item \textbf{Data Tiling} (as done by Ozturk et al. \cite{Ozturk_2009}): data tiling is applied to the original arrays, and any tile that is accessed is entirely transferred. The reported value corresponds to the best performing tile size that is less or equal to the iteration tile size.
\end{itemize}

The original layout baseline introduces no transfer redundancy but has the smallest amount of and the shortest burst transfers. The two other baselines are tradeoffs between burst usefulness and bandwidth use: using a bounding box or data tiling result in using only long burst accesses at the price of transfer redundancy.

Each benchmark is tested against a variety of tile sizes, with 1:1, 1.5:1 and 2:1 ratios.

\subsection{Results and discussion}

\begin{figure*}
\centering
	\begin{subfigure}{.48\textwidth}
		\centering
		\includegraphics[width=\columnwidth]{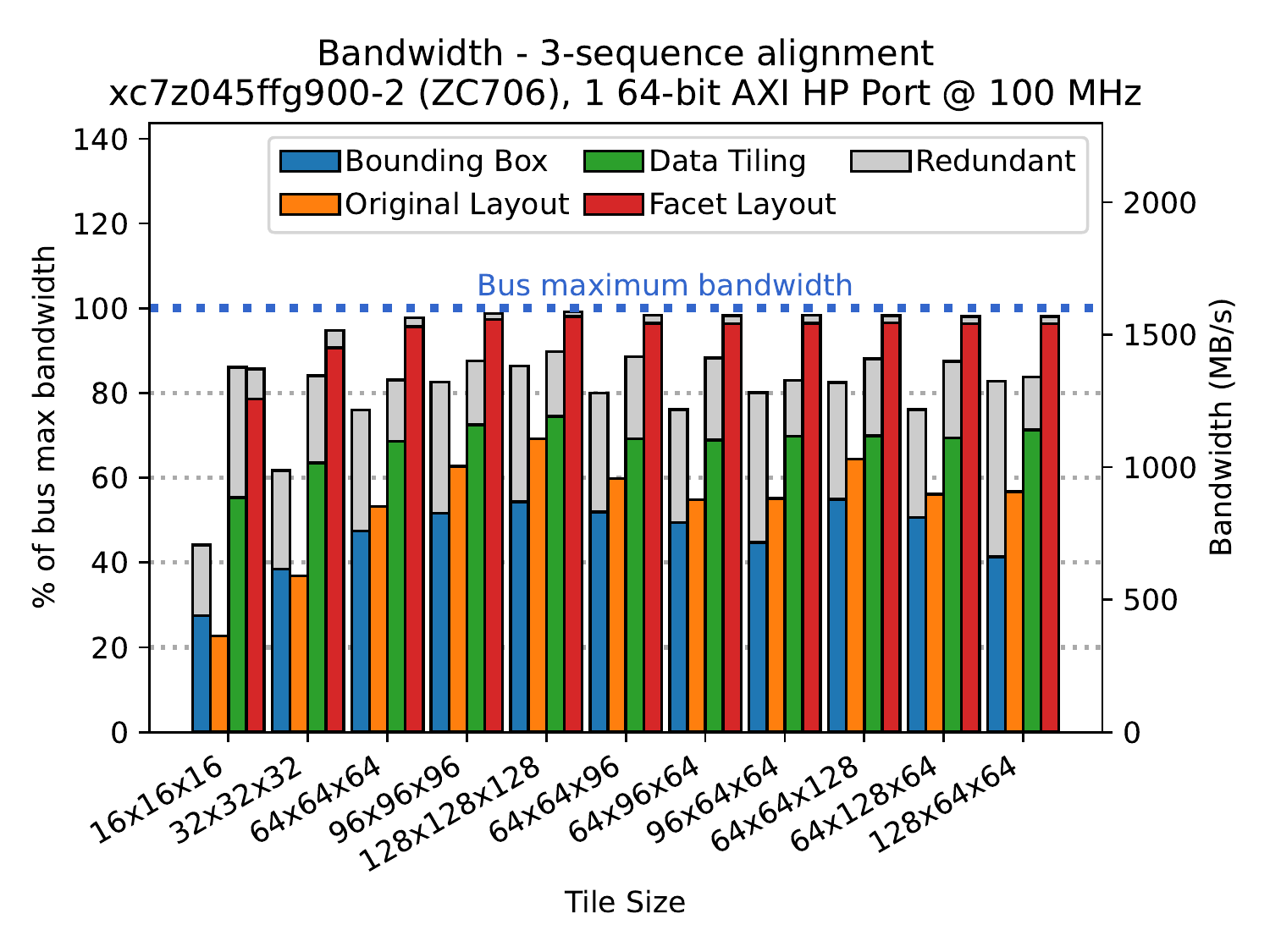}
		\caption{\texttt{sw3d}}
		\label{fig:sw3d-bandwidth}
	\end{subfigure}
	\begin{subfigure}{.48\textwidth}
		\centering
		\includegraphics[width=\columnwidth]{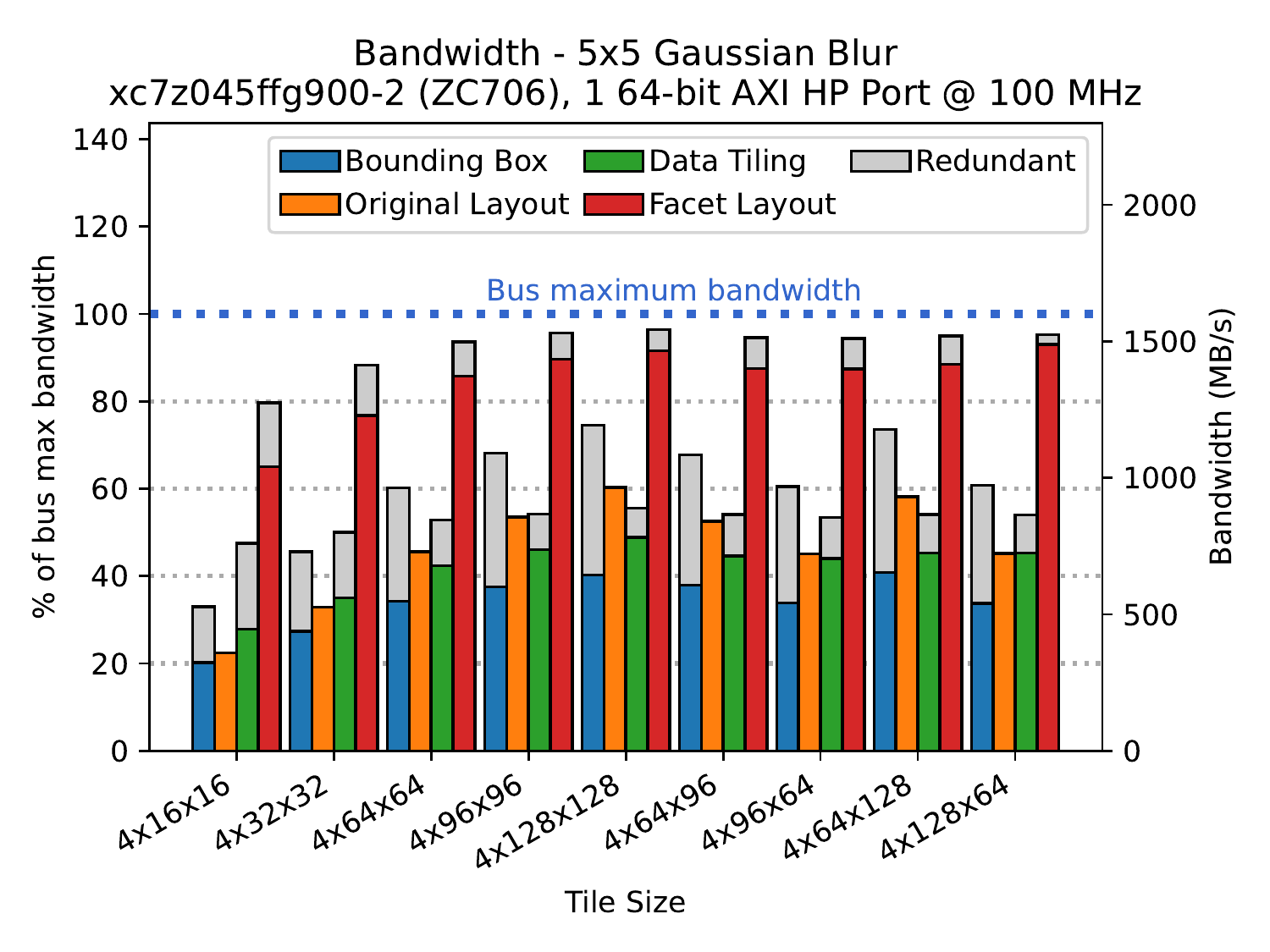}
		\caption{\texttt{gaussian}}
		\label{fig:gaussian-bandwidth}
	\end{subfigure}
	\begin{subfigure}{.48\textwidth}
		\centering
		\includegraphics[width=\columnwidth]{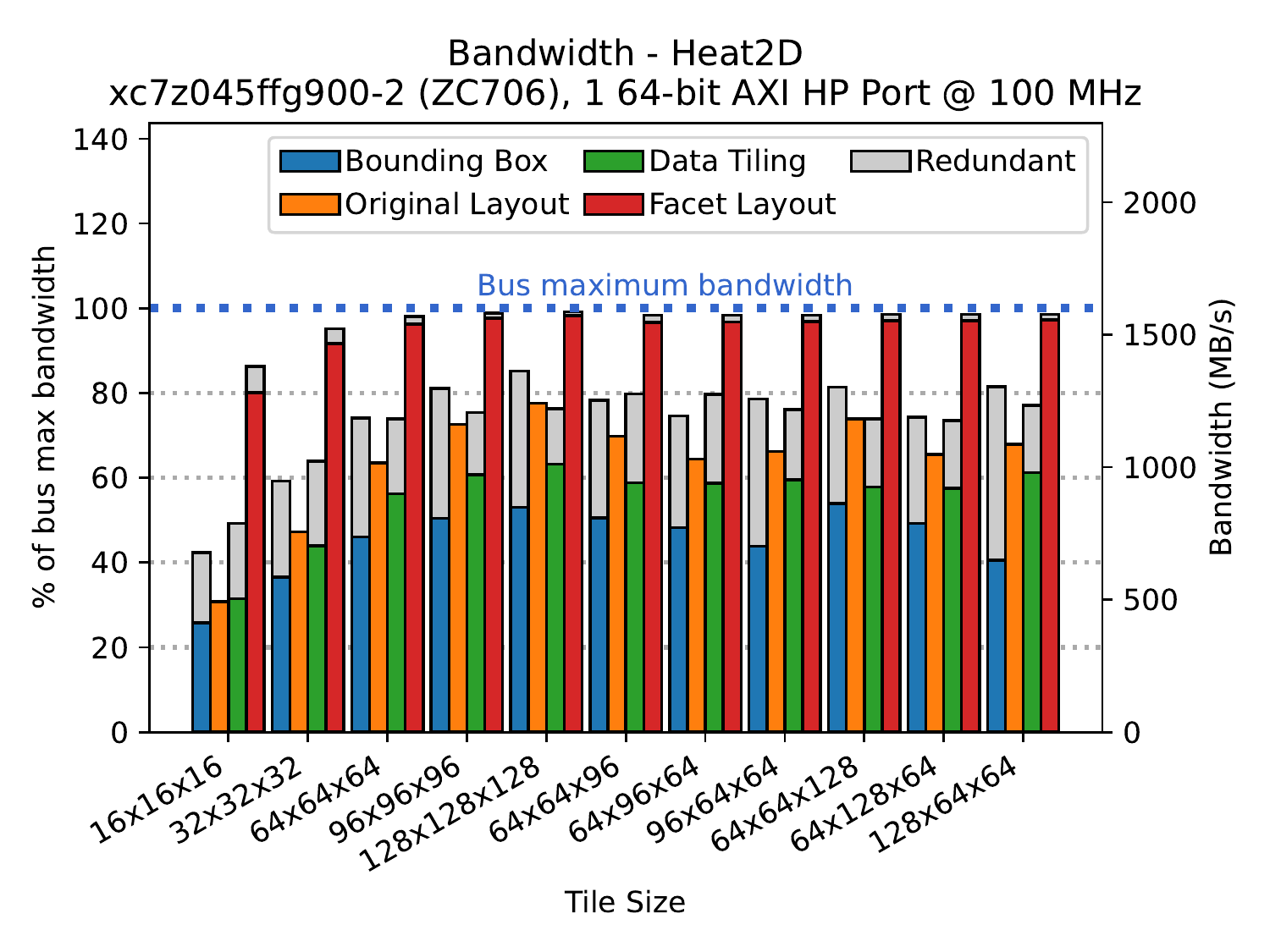}
		\caption{\texttt{jacobi2d5p}}
		\label{fig:jacobi2d5p-bandwidth}
	\end{subfigure}
	\begin{subfigure}{.48\textwidth}
		\centering
		\includegraphics[width=\columnwidth]{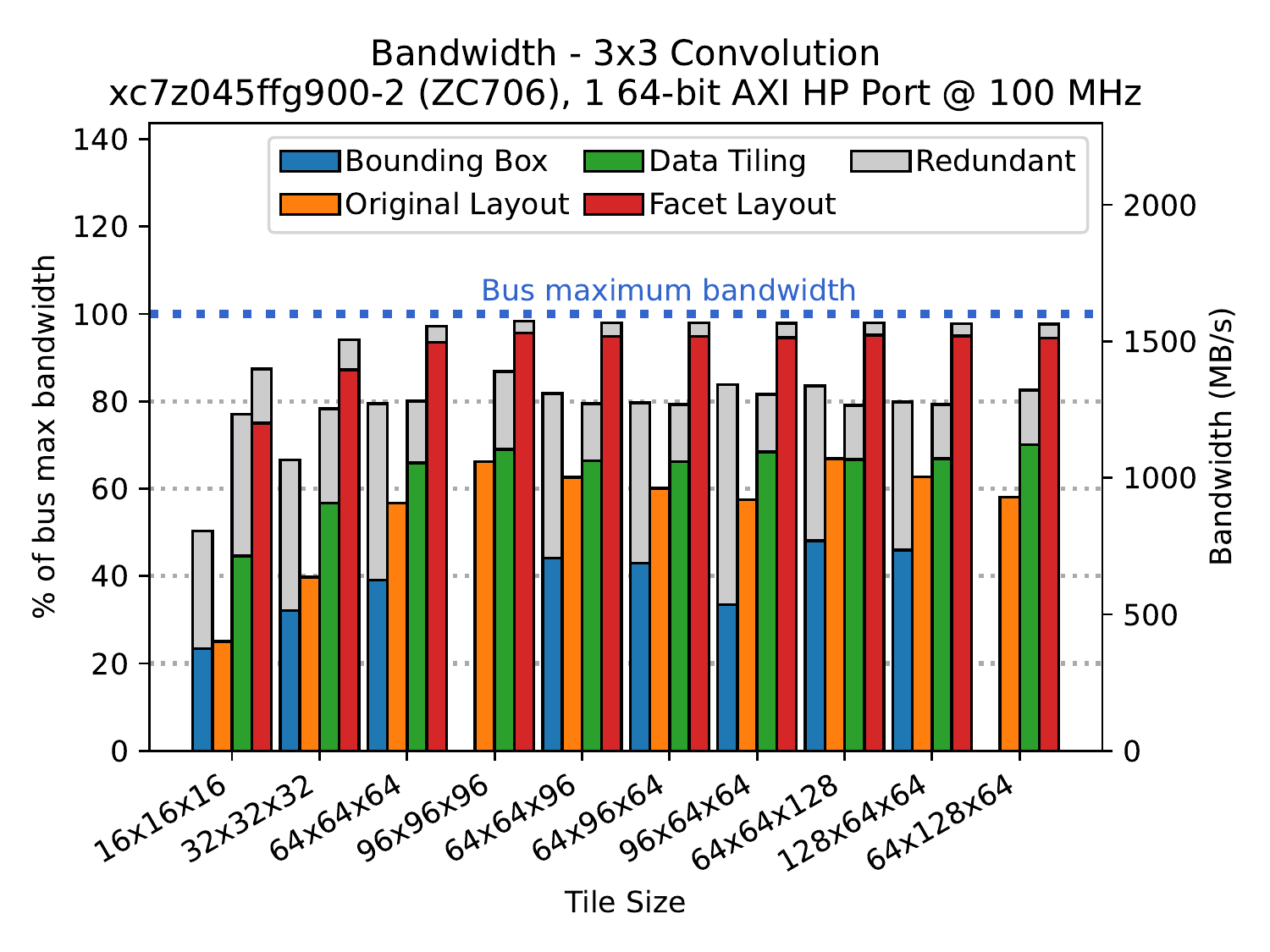}
		\caption{\texttt{jacobi2d9p}}
		\label{fig:jac2d9p-bandwidth}
	\end{subfigure}
	\begin{subfigure}{.48\textwidth}
		\centering
		\includegraphics[width=\columnwidth]{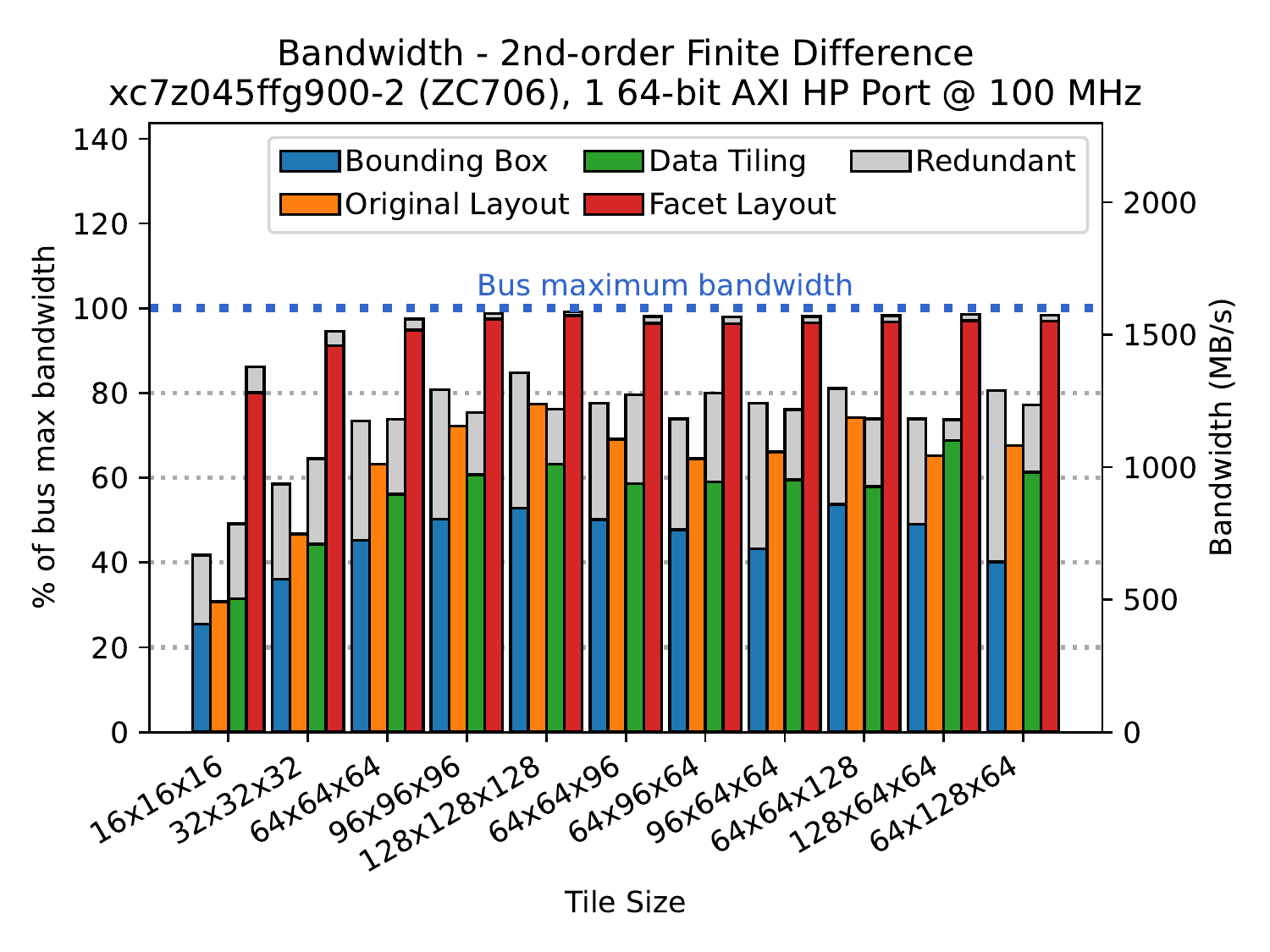}
		\caption{\texttt{jacobi2d9p-gol}}
		\label{fig:jac29pgol-bandwidth}
	\end{subfigure}
\caption{Bandwidth for all baselines, per benchmark. CFA is efficient even for tiles where one dimension is much smaller than the others. Bounding box is the technique used by Pouchet et al. \cite{Pouchet_2013}, Data Tiling is used by Ozturk et al. \cite{Ozturk_2009}, and the original layout is the best-effort baseline as in Bayliss et al. \cite{Bayliss_2012}.}
\label{fig:bandwidth}
\end{figure*}

\subsubsection{Raw bandwidth}
The raw bandwidth shown on Figure \ref{fig:bandwidth} indicates that the CFA layout and access pattern can reach close to 100\% of the bus' maximum bandwidth, whereas other baselines exhibit high redundancy overhead (especially with the bounding box).

The high efficiency of our approach is mainly explained by three points:
\begin{itemize}
    \item The small number of burst transfers per tile (4 in the case of 3-dimensional tiles). The data tiling approach uses a single burst access per tile, and the original layout and bounding box approaches issue multiple burst requests.
    \item The length of these burst transfers: one facet for the CFA approach,
    \item The ability of  Vitis HLS to use burst access overlapping, which hides latency for long bursts even when they are decomposed into smaller burst accesses.
\end{itemize}

\subsubsection{Effective bandwidth}
The effective bandwidth assesses the usefulness of the transferred data: it only counts the data transferred that is actually useful for the application. Data transferred then ignored is  consuming bus time, thus lowering the effective bandwidth. Figure \ref{fig:bandwidth} shows the effective bandwidth with colors, the difference with raw bandwidth being in grey. Two observations can be made:
\begin{itemize}
  \item For the considered tile sizes, CFA is able to bring the effective bandwidth close to 100\% of the bus bandwidth, which other allocations will not achieve.
  \item CFA is efficient even with small tile sizes. The \texttt{gaussian} benchmark, tiled with a small size in time ($4$) and larger spatial sizes (up to $128\times128$), shows that CFA exceeds 80\% of the bus bandwidth for tile sizes above $4\times64\times64$.
\end{itemize} 
The high usefulness of CFA (low difference between effective and raw bandwidth) is due to the choice of projections in CFA, which yields minimal redundancy. 

\subsubsection{Area cost}
We analyze two distinct cost metrics specific to FPGA designs: the computational resources (Slice and DSP), and the storage resources (Block RAM).

\paragraph{Computational resources}
\keymsg{The cost of CFA itself in terms of hardware is the address generators. These are small, as about 95\% of the logic area on our test platform remains available for compute engines.}
\begin{figure*}[ht]
\centering
	\begin{subfigure}{.48\textwidth}
		\centering
		\includegraphics[width=\columnwidth]{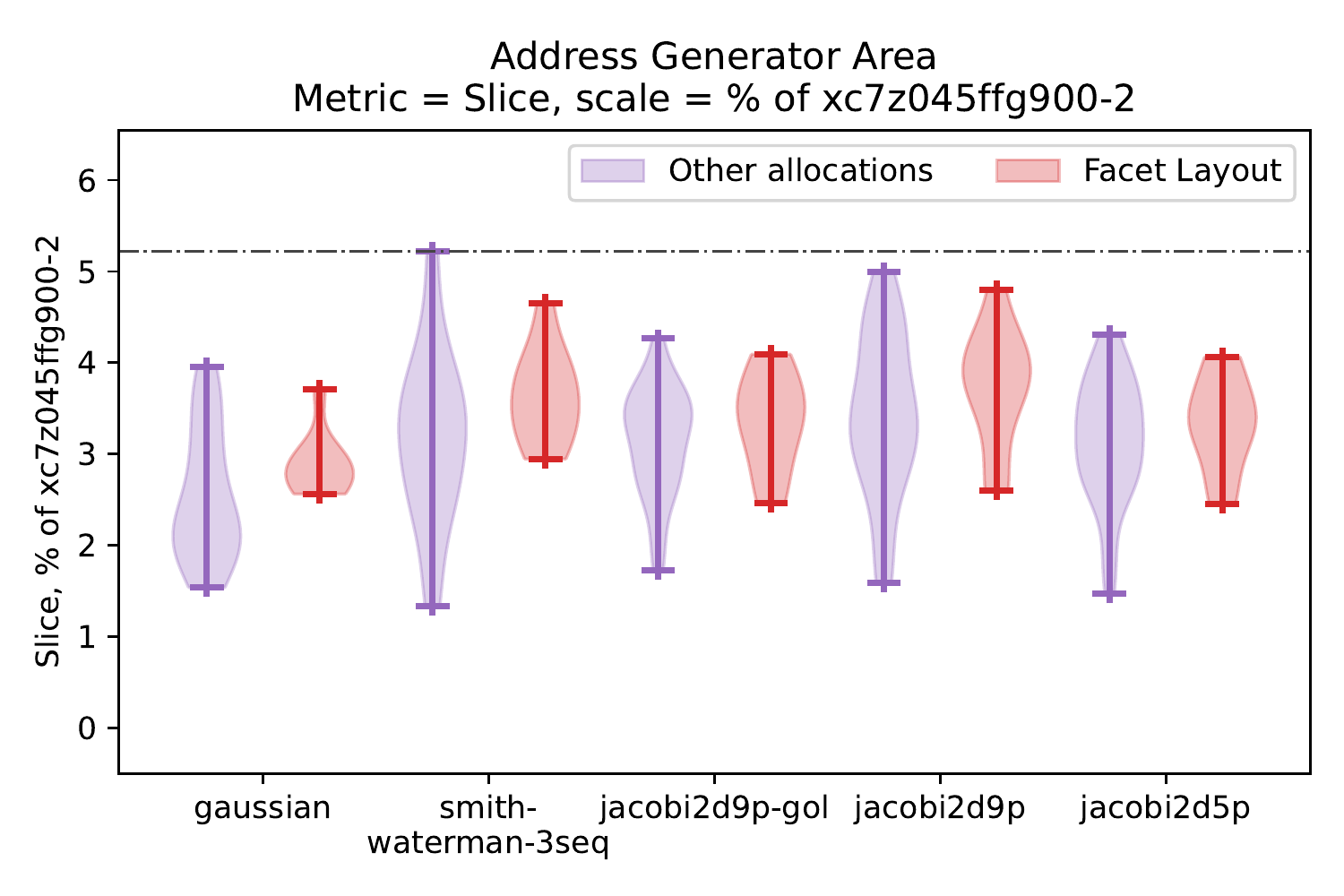}
		\caption{Occupancy of logic slices}
		\label{fig:area-Slice}
	\end{subfigure}
	\begin{subfigure}{.48\textwidth}
		\centering
		\includegraphics[width=\columnwidth]{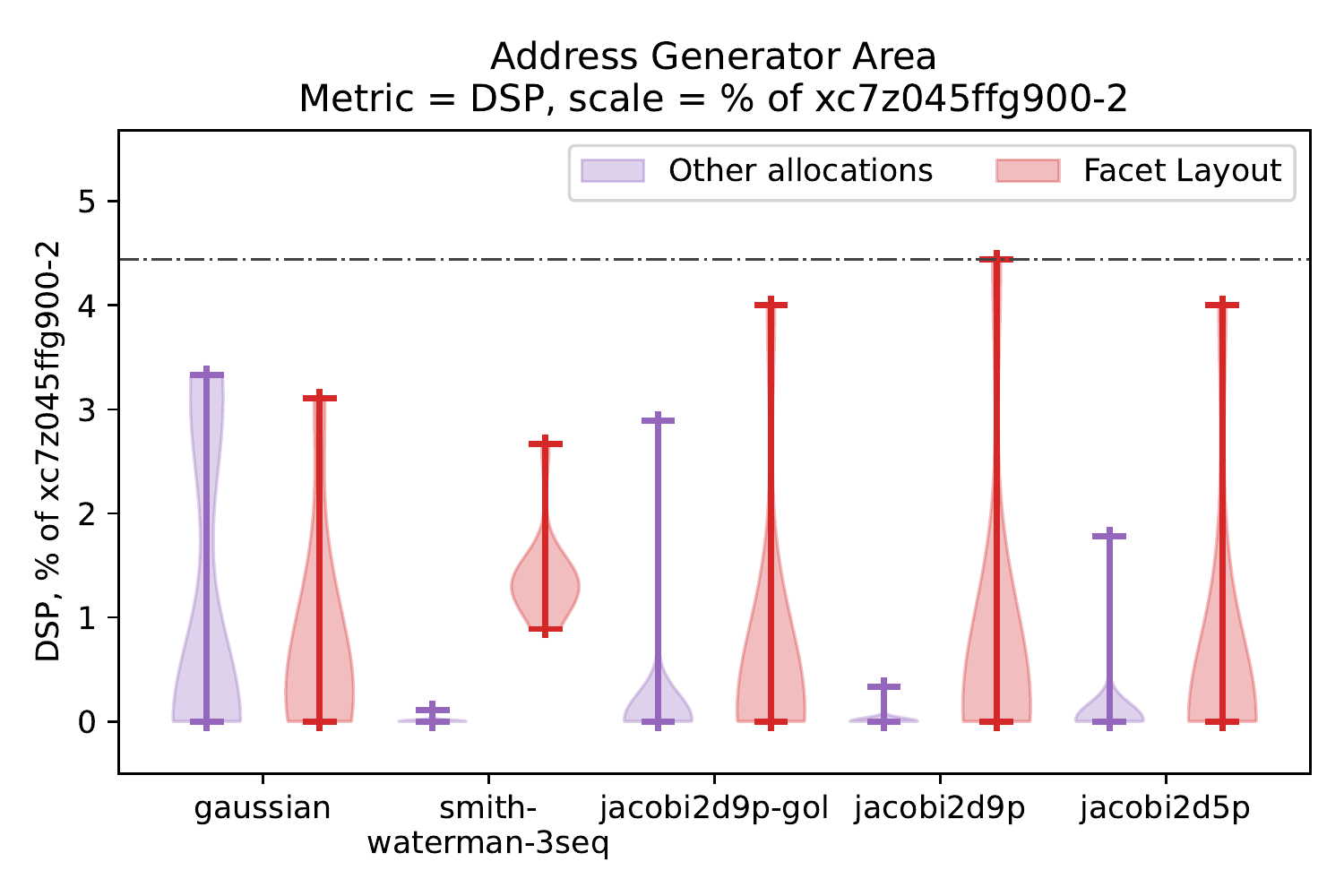}
		\caption{Occupancy of DSP blocks}
		\label{fig:area-DSP}
	\end{subfigure}
\caption{Area occupied by CFA and all other baselines (aggregated), as a percentage of the available area of xc7z045ffg900-2. The vertical lines span from minimum to maximum.}
\label{fig:area}
\end{figure*}

Regardless of the off-chip and on-chip allocations, the read and write engines take up a small fraction of the available logic resources. Figure \ref{fig:area} aggregates the area occupied by all baselines other than CFA for all tile sizes we have tried, and positions CFA. It can be observed that with tile sizes ranging from $16^3$ to $128^3$, except \texttt{gaussian} which tile sizes range from $4\times 16^2$ to $4\times 128^2$, designs occupy between $2$ and $5\%$ of the total slice area, and $0$ to $3\%$ of the total available DSP resources on an XC7Z045 FPGA chip. Canonical Facet Allocation does not show a significantly different slice occupancy than other baselines. For all benchmarks except \texttt{jacobi2d5p}, CFA requires some DSP blocks for some tile sizes, which are used to compute off-chip base addresses, but never needs more than 40 out of 900 (4\%) of the DSP resources. The same observation holds for the \texttt{gaussian} benchmark, where the baseline using the original allocation takes between 26 and 30 DSP blocks.

Two conclusions can be drawn:
\begin{itemize}
\item Address generators are small units in terms of area, regardless of the allocation,
\item CFA does not exhibit a different area pattern than 
\end{itemize}

It should additionnally be noted that the experiments were carried out without any compute logic on the accelerators; given the small size of the memory access modules, the synthesis tool did not have to significantly optimize the design for area.

\paragraph{Block RAM usage}
\keymsg{Using CFA does not change the on-chip allocation, therefore using CFA does not significantly increase the BRAM cost of a design.}

\begin{figure}
\centering
\includegraphics[width=\columnwidth]{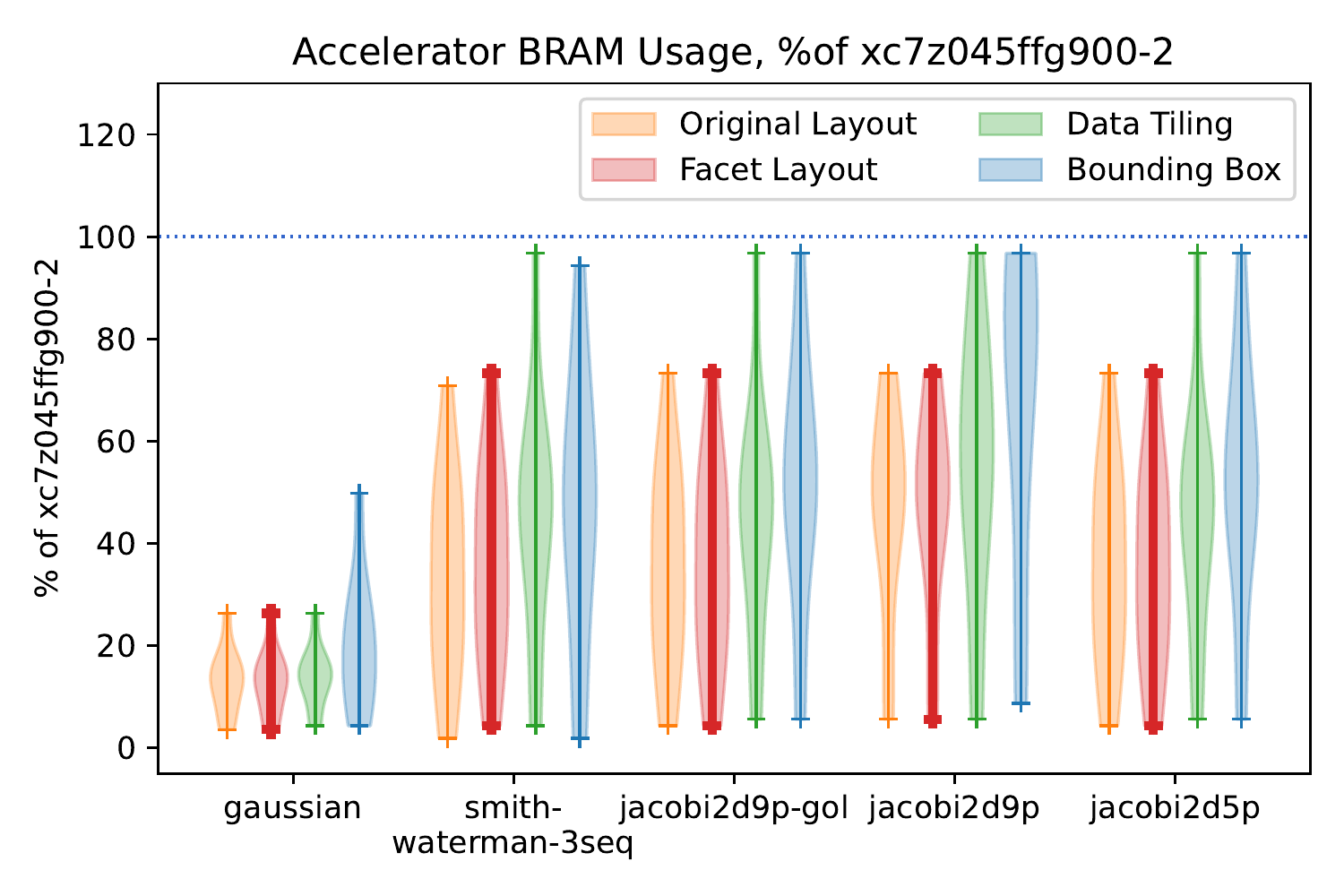}
\caption{BRAMs occupied by all baselines (in color), for each benchmark, as a proportion of BRAMs available in xc7z045ffg900-2.}
\label{fig:bram}
\end{figure}

In an FPGA design, Block RAM resources used for on-chip data storage are shared between multiple actors. Even when the compute actor is not implemented, the memories needed to hold all the data on chip in and out of the memory actor must be present. Therefore, all designs do occupy a significant proportion, up to 95\% of the available on-chip Block RAM, as Figure \ref{fig:bram} shows. BRAM was, indeed, the factor limiting tile size - the larger the tile, the more data needs to fit into on-chip memories. 

We can observe on Figure \ref{fig:bram} that the distribution of on-chip memories using CFA and the original allocation is the same, with an exception (\texttt{smith-waterman-3seq}). This is due to the fact that CFA does not change the on-chip allocation, which is defining the amount of on-chip memory needed. The BRAM overhead for bounding box and data tiling baselines is mainly due to two facts: for the bounding box baseline, writing a superset of the tile footprint implies that the values written while not modified have been read and held on chip.

\section{Conclusion}

Our work provides an answer to the under-utilization of memory bandwidth observed in many instances where an FPGA or ASIC accelerator is developed. The insufficient observed memory bandwidth results in stalls, preventing the full exploitation of the on-chip parallelism.

Memory allocation can be the cause of a significant under-utilization of the available bandwidth, due to the high latency of element-wise accesses. By introducing a data layout in memory for programs with uniform dependencies that is burst-friendly, we are able to overcome the under-utilization of the memory bandwidth and use it to its full extent. The method can be automated: we have developed a proof-of-concept procedure and compiler pass that implements the main code generation feature of CFA, which is the on-chip and off-chip address generator.

To further increase the benefits of CFA, the machine model we have considered may be extended to multi-port memory accesses, such as high-bandwidth memory, and distributed memories. In such architectures, there are multiple data ports; to benefit from all their bandwidth, one has to find an adequate repartition of data over each memory port to balance accesses. 

\bibliography{main}
\bibliographystyle{ieeetr}

\cleardoublepage

\appendix

\subsection{Set-wise construction of facets}

Reminder of the notations:
\begin{itemize}
\item $E\subset\mathrm{vect}\left(\vec{e_{1}},\dots,\vec{e_{d}}\right)$
: Iteration space (rectangular subspace of $\mathbf{Z}^{d}$), with
$\left(\vec{e_{1}},\dots,\vec{e_{d}}\right)$ an orthonormal base
\[
E=\left\{ \vec{x}=\left(x_{1},\dots,x_{d}\right):0\leqslant x_{1}<N_{1},\dots,0\leqslant x_{d}<N_{d}\right\} 
\]
\item $d$ : number of dimensions of $E$
\item $\vec{B_1}, \dots, \vec{B_p}$ : dependence vectors, such that rectangular tiling is legal. We assume all dependence vectors are backwards in all dimensions: $\forall i, j: \vec{B_i}\cdot\vec{e_j} \leqslant 0$.
\item $N_{1},\dots,N_{d}$ : iteration space size (assuming a rectangular
iteration space)
\item $t_{1},\dots,t_{d}\in\mathbf{N}$ : tile sizes (assume $N_i$ if no tiling on dimension $i$)
\end{itemize}

Let $k \in \intset{1,d}$, $\vec{e_{k}}$ the canonical vector of the $k$-th dimension. The $k$-th \emph{face} is given by those iterations which $k$-th coordinate is equal to $t_{k}-1$.

The depth of the $k$-th facet is:
$$w_{k}=\max_{q\in \intset{1,p} } \left|p_{\vec{e_{k}}}\left(\vec{B_{q}}\right)\right| = \max_{q\in \intset{1,p} } \left|\vec{e_{k}} \cdot \vec{B_{q}}\right|$$

Let a tile of iterations be 
$$ T_{i_1, \dots, i_d} = \left\lbrace \vec{x}=\left(x_{1},\dots,x_{d}\right):\forall q\in\intset{1,d} :\left\lfloor \frac{x_{q}}{t_{q}}\right\rfloor =i_{q} \right\rbrace $$

The $k$-th facet for tile $T$ is the set of iterations given by:
$$S_{k}(T)=\left\{\left(x_{1},\dots,x_{d}\right)\in T : t_{k}-w_{k} \leqslant x_k \mod t_k\right\}$$

\subsection{Proof that the flow-out of every tile is contained inside facets}

In order to replace the existing off-chip memory accesses by facet accesses, all the flow-in iterations of a tile need to be contained inside facets. 

\begin{proposition}
Flow-in iterations are contained inside facets.

\textbf{Purpose:} Proves that all the flow-in data can be retrieved from facets.
\begin{IEEEproof}

The iteration-wise flow-in set of a tile is defined as those iterations which result is used by this tile but are executed in another tile. It can be written as:

$$ \varphi_i(T) = \left\lbrace \vec{y} \in E\setminus T : \exists j \in \intset{1,p} : \vec{y} - \vec{B_{j}} \in T \right\rbrace$$

Let $T = T_{i_1, \dots, i_d}$ be a tile of iterations, $\vec{y} = \left(y_1,\dots,y_d\right) \in \varphi_i(T)$, and $T'$ be the tile of iterations containing $\vec{y}$. Let $j\in \intset{1,p}$ such that $\vec{y} - \vec{B_{j}} \in T$.
Obviously, $\vec{B_j}$ is non-null, so let $q \in \intset{1,d}$ such that $\vec{B_j}\cdot\vec{e_q} < 0$. 

Let's first show that there exists a $q$ such that $\vec{B_j}\cdot\vec{e_q} \neq 0$ and $\left\lfloor\frac{y_q}{t_q} \right\rfloor \neq i_q$: given that $\vec{B_j} \neq \vec{0}$, there exists $q$ such that $\vec{B_j}\cdot\vec{e_q} \neq 0$; if any such $q$ verifies $\left\lfloor\frac{y_q}{t_q} \right\rfloor = i_q$, then for all $r \in \left\lbrace 1,\dots,d \right\rbrace$, $\left\lfloor\frac{y_r}{t_r} \right\rfloor = i_r$, thus $\vec{y} \in T$, which is false. 

Let $q$ be such that $\vec{B_j}\cdot\vec{e_q} \neq 0$ and $\left\lfloor\frac{y_q}{t_q} \right\rfloor \neq i_q$. By definition, $w_q \geqslant \left| \vec{B_j}\cdot\vec{e_q} \right|$. We have: 

$$
\begin{array}{lcr}
\left\lfloor\frac{y_q}{t_q} \right\rfloor \neq i_q & \text{ and } & \left\lfloor \frac{y_q - \vec{B_j}\cdot\vec{e_q}}{t_q} \right\rfloor = i_q
\end{array}
$$

As $\vec{B_j}\cdot\vec{e_q} < 0$, $\left\lfloor\frac{y_q}{t_q} \right\rfloor < i_q$. We thus have $$ y_q \mod t_q - \vec{B_j}\cdot\vec{e_q} \geqslant t_q $$
and as $w_q \geqslant \vec{B_j}\cdot\vec{e_q}$, $$ y_q \mod t_q \geqslant t_q - w_q$$

which proves that $\vec{y} \in S_q(T')$. As a result, \emph{the iteration flow-in of a tile is contained inside an union of facets}.

\end{IEEEproof}
\end{proposition}

\end{document}